\documentclass[12pt]{article}
\usepackage{amssymb, amsmath,booktabs,bm}
\usepackage{graphicx}
\graphicspath{{figures/}}
\usepackage{array,tabularx}
\usepackage{float}
\usepackage{microtype}
\usepackage{xcolor}
\usepackage{hyperref}
\hypersetup{hidelinks}

\usepackage{authblk}
\usepackage[round,authoryear]{natbib}

\newcommand{\MVN}{\mathcal{N}}
\newcommand{\insertplaceholder}[2]{%
  \begin{center}\vspace{0.35em}\textit{[Insert #1~\ref{#2} here]}\vspace{0.35em}\end{center}%
}

\title{\bf Bayesian Spatio-temporal Modeling of Multi-Site Wastewater Surveillance Data}
\author[a,c]{K.\ Ken Peng}
\author[b]{Chengqian Xian\thanks{Corresponding author. Chengqian Xian,
School of Science,
Harbin Institute of Technology, Shenzhen,
HIT Campus of University Town of Shenzhen,
Shenzhen 518055, China.
Email: \href{mailto:xianchengqian@hit.edu.cn}
{xianchengqian@hit.edu.cn}.}}
\author[d]{Charmaine B.\ Dean}
\author[a]{Robert\ Delatolla}
\author[c]{X.\ Joan Hu}
\author[a]{Elizabeth M.\ Renouf}

\affil[a]{Department of Civil Engineering, University of Ottawa\\

75 Laurier Avenue East, Ottawa, ON, K1N 6N5, Canada}

\affil[b]{School of Science, Harbin Institute of Technology, Shenzhen, China}

\affil[c]{Department of Statistics and Actuarial Science, Simon Fraser University\\
8888 University Dr, Burnaby, BC, V5A 1S6, Canada}

\affil[d]{Department of Statistics and Actuarial Science, University of Waterloo\\
200 University Ave W, Waterloo, ON, N2L 3G1, Canada}

\date{}

\begin{document}

\maketitle

\begin{abstract}
Wastewater surveillance provides population-level evidence of infectious disease activity. As surveillance expands across multiple sites, repeated measurements must be analyzed jointly while accounting for differences among sites and the public health regions they represent. We develop a Bayesian spatio-temporal framework that decomposes wastewater measurements into a shared epidemic curve, spatially correlated regional deviations, persistent site effects, and within-site temporal dependence. This hierarchical decomposition accommodates unequal observation windows and sampling schedules, allowing regional trajectories to be compared while preserving site-level heterogeneity. A direct marginal spatial covariance specification further separates the magnitude of regional variation from the strength of association between neighboring regions, providing distinct interpretations of these two sources of spatial variation. We also introduce region-specific time shifts to distinguish regional leads and lags in epidemic timing from differences in signal levels and local fluctuations. Simulation studies demonstrate the recovery of these components and show that omitting true time shifts can lead to increased estimates of residual temporal variation. Application to Ontario SARS-CoV-2 wastewater surveillance data reveals persistent site heterogeneity, within-site temporal dependence, and spatial association among neighboring regions. Regional timing differences are more pronounced during the first year of Omicron than in the XBB-and-later period.
\end{abstract}

\section{Introduction}
\label{sec:introduction}

Wastewater and environmental surveillance has become an important population-level tool for monitoring infectious disease activity. By measuring biological and chemical markers in wastewater, surveillance systems can provide disease prevalence information on a community without requiring individual-level testing. Applications include monitoring substances associated with population-level drug use and detecting pathogens \citep{asghar2014environmental, zuccato2008estimating}. During the COVID-19 pandemic, wastewater measurements of SARS-CoV-2 RNA were shown to track community infection dynamics and to complement clinical surveillance, especially when individual testing was incomplete or biased \citep{medema2020presence,peccia2020measurement,randazzo2020sars,larsen2020tracking}. More recent reviews emphasize that wastewater surveillance is now being considered as a broader public health infrastructure for SARS-CoV-2 and beyond \citep{parkins2024wastewater}.

The statistical literature on wastewater surveillance has developed rapidly, addressing preprocessing, temporal smoothing, forecasting, and epidemiological interpretation of wastewater concentrations \citep{arabzadeh2021data,dai2022statistical,schoen2022sars,galani2022sars}. Other work relates wastewater signals to clinical outcomes through Bayesian functional analysis \citep{dai2024bayesian}, distributed-lag and Markov-modulated analyses \citep{peng2023exploration,PENG2025100840}, and joint modeling through a latent infection process \citep{peng2026joint}. For surveillance programs with multiple sites, inference also requires a model for how their repeated measurements relate across sites and regions.

The present study is motivated by SARS-CoV-2 wastewater surveillance data collected through Ontario's Wastewater Surveillance Initiative, a provincial sampling network across public health unit regions. Public Health Ontario, the provincial public health agency, reported wastewater trends from this network as a complement to clinical surveillance \citep{pho2023wastewater}. The analyzed data contain repeated measurements from 56 wastewater sampling sites distributed across 33 public health units (PHUs), covering the period from July 2020 to July 2024. This surveillance network provides an opportunity to investigate regional epidemic patterns while accounting for differences among sampling sites and their temporal observation histories.

Multi-site wastewater surveillance data are not a balanced spatio-temporal array, since observation windows differ across sites, and sampling frequency can vary over time. Site-specific catchment characteristics also create persistent differences in signal level. Analyzing sites independently ignores shared regional information, whereas averaging sites within regions can obscure site heterogeneity. Joint analysis places regional trajectories on a common temporal reference while retaining site differences and sharing information across regions. Spatio-temporal models provide tools for representing these dependencies \citep{cressie2011statistics}, but their components must be matched to the sampling hierarchy and to the available information.

Several recent papers model wastewater signals across locations. \citet{li2023wastewater} combine region and sampling-site effects, a common temporal component, and a continuous-space spatio-temporal process based on treatment-plant locations. \citet{somerset2025differentiable} use differentiable Gaussian processes to estimate wastewater signals and their rates of change from irregular and censored observations. \citet{sun2025dynamic} jointly model wastewater and clinical trajectories, allowing spatial dependence in the clinical component. Related spatial functional models have been developed for infectious-disease counts \citep{ma2026spatial}. These approaches show the value of spatial and temporal modeling, but leave less developed a wastewater model that simultaneously distinguishes regional spatial variation, persistent differences among sites within regions, and within-site temporal residual dependence. That decomposition is the focus here.

Each site is mapped by its sampling location, but its measurement reflects upstream contributions from a sewershed whose boundaries may not be fully known. We therefore retain individual sites within public health units and use regional adjacency to describe broad spatial association. This spatial component quantifies how neighboring PHUs vary together beyond the shared temporal trajectory and persistent site differences. A direct marginal covariance specification makes the regional variance and neighboring association explicit \citep{wall2004close}. It allows the magnitude of regional variation to be distinguished from the strength of local association within the hierarchical variance decomposition. We assess parameter recovery under the intended surveillance design.

The resulting Bayesian framework has three main contributions. First, a hierarchical decomposition places regional trajectories on a common statistical basis despite unequal site histories. It retains persistent site differences while separating shared temporal change, regional fluctuations, and within-site temporal dependence, so regional comparisons distinguish these sources of variation. Second, a direct marginal spatial covariance distinguishes the magnitude of regional fluctuations from neighboring association, quantifying both how much regions depart from the shared trajectory and how those departures vary together. Third, region-specific time shifts, adapted from curve registration \citep{telesca2008registration}, separate regional leads and lags from differences in signal level and local fluctuations, allowing epidemic timing to be compared across regions and periods. Simulations assess recovery of these components and the effect of ignoring time shifts. The Ontario case study shows how site heterogeneity, neighboring regional association, and timing differences contribute to the surveillance patterns.

The remainder of the paper is organized as follows. Section~\ref{sec:data} describes the Ontario data and the motivating statistical challenges. Section~\ref{sec:model} presents the baseline framework, spatial specification, regional time-shift extension, and estimation procedure. Section~\ref{sec:case-study} applies both baseline model and model with time shift extension to the Ontario wastewater data. Section~\ref{sec:simulation} uses Ontario-motivated simulations to assess parameter recovery and regional timing. Section~\ref{sec:discussion} discusses interpretation and future investigations.

\section{Ontario Wastewater Surveillance Data}
\label{sec:data}

\subsection{Study setting and data overview}
\label{sec:data-study-setting}


The wastewater viral load measurement is derived from the normalized N1/N2 wastewater signal. This signal is based on the SARS-CoV-2 N1 and N2 gene targets and is normalized by the concentration of pepper mild mottle virus (PMMoV), an indicator of human fecal content commonly used to adjust for variation in wastewater dilution and sample concentration. Data preprocessing first retained records from wastewater treatment plant sewersheds and excluded backup or inactive sewershed types. The source file covers 34 public health units, but York Region Public Health has only upstream sampling sites and was excluded by this restriction. This yielded 56 wastewater sites across 33 public health units, contributing 60,165 dated rows in the raw file, of which 29,295 had non-missing viral signal measurements.

\begin{figure}[H]
\centering
\includegraphics[width=0.72\textwidth]{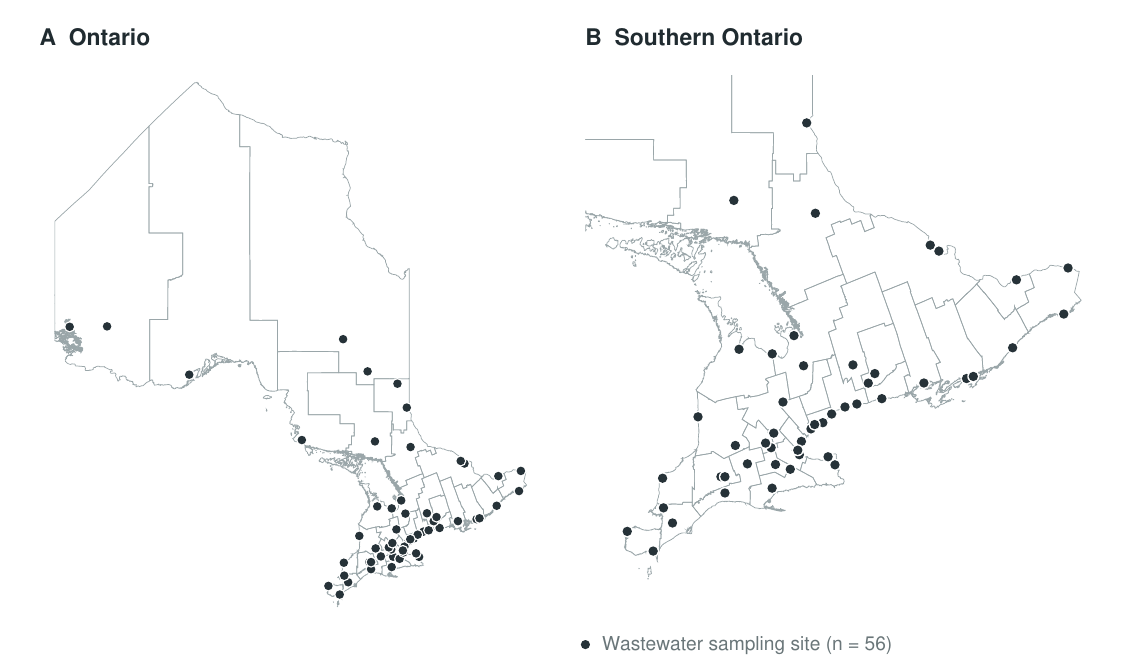}
\caption{Spatial structure for the 33 public health units used in the Ontario wastewater analysis. Edges indicate neighboring public health units in the adjacency graph. Points are the wastewater sites.}
\label{fig:data-spatial-graph}
\end{figure}

The primary spatial resolution considered in this manuscript is the 33 public health unit structure. Figure~\ref{fig:data-spatial-graph} shows the adjacency graph used to describe neighboring public health units. The spatial structure is irregular. Some units have many neighbors, whereas others have few. The 56 wastewater sites are also unevenly distributed across the 33 units, so the amount of site-level information differs across regions. This uneven coverage spans densely sampled urban catchments and larger units represented by fewer sites.

\insertplaceholder{Figure}{fig:data-spatial-graph}

For all descriptive figures and model-based analyses, the site-level weekly arithmetic mean signal is transformed using the natural logarithm. This reduces the influence of very large weekly values and represents proportional differences in viral signal as additive differences, facilitating comparison of epidemic trajectories across sites and periods.

Sampling dates start from July 14, 2020, and we consider the data up to July 29, 2024. The time axis was aligned to complete seven-day intervals, with week 1 beginning on July 14, 2020; the final complete week ended on July 29, 2024, giving 211 study weeks. Within each site and week, available normalized N1/N2 measurements were first averaged arithmetically on their original scale. The natural logarithm of this weekly mean then formed the site-week observation. The common seven-day intervals provide a consistent reporting scale for comparing regional rises and falls and expressing time shifts in weeks. Weekly summaries emphasize sustained epidemic patterns over short-term fluctuations; restricting the record to complete weeks also avoids comparing a partial final interval with full weeks. This aggregation is a choice for interpretation of the Ontario surveillance data, not a requirement of the model, which accommodates irregular observation times and unequal observation windows. Within the recorded site-specific observation windows, weekly aggregation yielded 8,646 site-week cells, of which 8,562 contained a non-missing measurement. The resulting measurements are noisy population-level indicators for the wastewater catchments served by the sampling sites. The next two subsections describe the spatial organization of the sites and the main descriptive patterns in the observed weekly data.





\subsection{Spatial and temporal patterns}
\label{sec:data-patterns}

The location of a wastewater sampling site does not necessarily represent the population contributing to its measurements. Several sites may serve populations within the same public health unit and share regional influences. Because precise catchment boundaries are not consistently available, public-health-unit adjacency provides a practical basis for describing spatial relationships. Where catchment geography or population coverage is known, distance-based spatial relationships could also be considered.

\begin{figure}[H]
\centering
\includegraphics[width=\textwidth]{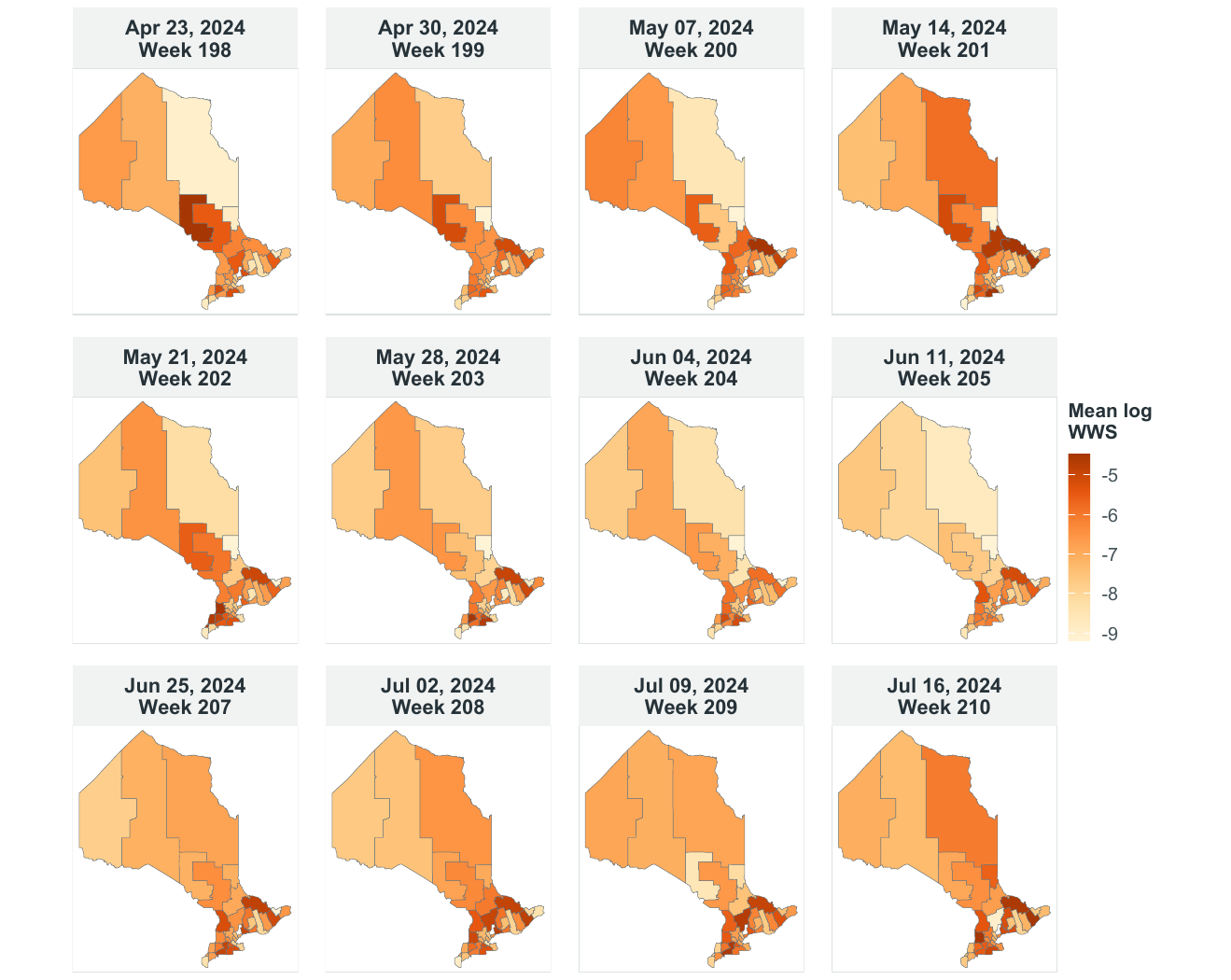}
\caption{Observed average log wastewater signal by public health unit over selected time periods.}
\label{fig:data-observed-map}
\end{figure}

Figure~\ref{fig:data-observed-map} gives a map-based summary of observed wastewater levels over selected time periods. Each panel represents one study week, and each row contains four consecutive weeks to illustrate the observed surveillance period. The color scale shows the public-health-unit average of the site-level weekly wastewater signal. 

Several features are observed from this figure. First, the wastewater signal is not spatially uniform. During high-signal periods, some public health units show much stronger wastewater levels than others. Second, the spatial pattern evolves gradually across consecutive weeks, suggesting temporal continuity rather than independent week-to-week variation. Third, neighboring public health units often have similar color intensity within a given week, suggesting spatial dependence, but the strength and location of this similarity vary over time. These features motivate a model that considers temporal evolution and spatial dependence jointly.

\insertplaceholder{Figure}{fig:data-observed-map}

\begin{figure}[H]
\centering
\includegraphics[width=\textwidth,height=.72\textheight,keepaspectratio]{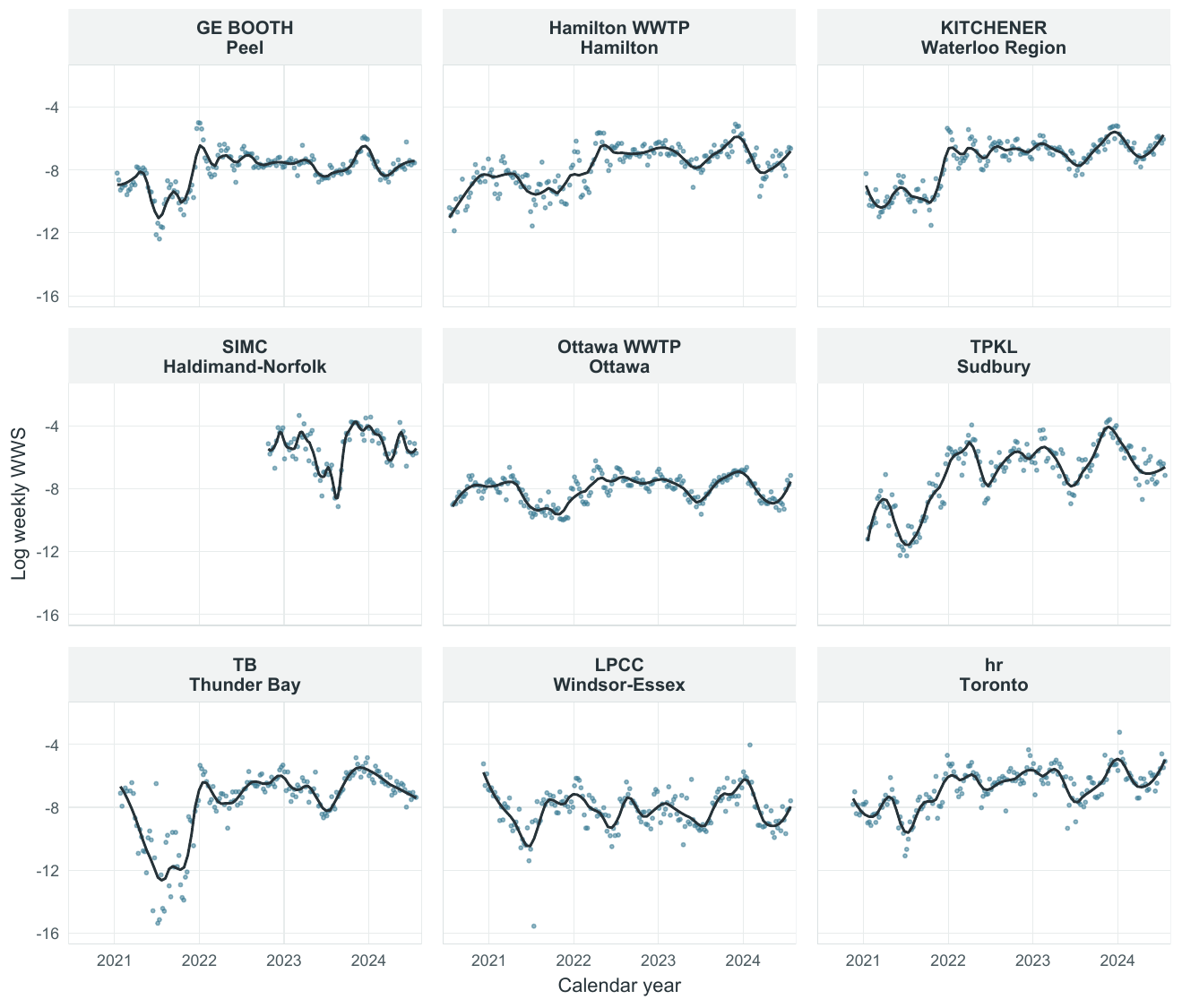}
\caption{Observed weekly log wastewater signals at nine selected sites. Lines are descriptive LOESS smooths, not model-based estimates. The same sites appear in Figure~\ref{fig:fitted-site-trajectories}; all 56 sites are shown in Web Figures 4--7.}
\label{fig:data-site-series}
\end{figure}

Figure~\ref{fig:data-site-series} shows nine representative site-level weekly trajectories. Web Appendix C and Web Figures 4--7 show all 56 and confirm that the broad epidemic waves coexist with differences in observation window and signal level. Although weekly aggregation produces relatively dense trajectories with limited missingness once a site is active, the signals remain noisy. These features motivate a decomposition that includes a shared temporal curve, regional deviations, persistent site-level differences, and within-site temporal residual dependence.

\insertplaceholder{Figure}{fig:data-site-series}

The broad waves also raise a question about regional timing. Do wastewater signals rise and fall together across Ontario, or do some regions lead and others lag? Because epidemic patterns can change over time, we examine this question separately in two Omicron periods. Section~\ref{sec:model-delay} introduces the model used to estimate these timing differences.

\section{Bayesian Spatio-temporal Model}
\label{sec:model}

The descriptive analysis motivates a model that pools information across the surveillance system while preserving regional and site-level variation. We first present the baseline framework, then specify its regional spatial covariance, add a regional time-shift extension, and describe a common Bayesian estimation procedure.

\subsection{Notation and modeling framework}
\label{sec:model-notation}

Let $i=1,\ldots,K$ index regions and let $j=1,\ldots,n_i$ index wastewater sites within region $i$. For example, in the Ontario motivating data, the regions are the 33 public health units described in Section~\ref{sec:data}. Let $\mathcal T_{ij}=\{t_{ij1},\ldots,t_{ijT_{ij}}\}$ be the sampling times for site $j$ in region $i$, where $T_{ij}$ is allowed to differ across sites. Let $S=\sum_{i=1}^K n_i$ denote the total number of sites and $N=\sum_{i=1}^K\sum_{j=1}^{n_i}T_{ij}$ denote the total number of observations.

Let $Y_{ij}(t)$ denote a wastewater surveillance measurement on the analysis scale at site $j$ in region $i$ at time $t$; in the Ontario analysis, time is measured in weeks and the response is the natural logarithm of the within-site arithmetic weekly mean of normalized N1/N2 measurements. At the sampling times for site $(i,j)$, let
$\bm Y_{ij}=(Y_{ij}(t_{ij1}),\ldots,Y_{ij}(t_{ijT_{ij}}))^{\top}$ and
$\bm e_{ij}=(e_{ij}(t_{ij1}),\ldots,e_{ij}(t_{ijT_{ij}}))^{\top}$.
We write the model at a generic time $t$; the observation equation is evaluated
only for $t\in\mathcal T_{ij}$ at site $(i,j)$. Model~1 (the baseline model) is
\begin{equation}
\begin{aligned}
Y_{ij}(t)
&=
g(t)+v_i(t)+u_{ij}+e_{ij}(t),\\
g(t)
&=
\sum_{m=1}^{M} B_m(t)\phi_m,\\
\bm v(t)
&=
(v_1(t),\ldots,v_K(t))^{\top}
\sim
\MVN\left(\bm 0,\Sigma_v\right),\\
u_{ij}
&\overset{\mathrm{iid}}{\sim}N(0,\sigma_u^2),\\
\operatorname{Cov}\{e_{ij}(t),e_{ij}(s)\}
&=\sigma_e^2\exp\{-|t-s|/\delta_e\}.
\end{aligned}
\label{eq:full-model}
\end{equation}
Here and throughout, $\MVN$ denotes a multivariate normal distribution. The regional process $\bm v(t)$ represents time-varying deviations from the shared curve. Its covariance $\Sigma_v$ is specified in Section~\ref{sec:spatial-formulations}. Each $e_{ij}(t)$ is a mean-zero Gaussian residual process with the covariance given in \eqref{eq:full-model}.

The shared temporal component $g(t)$ captures the broad wastewater trend common to all sites. We use cubic B-spline basis functions $B_m(t)$ and estimate their coefficients $\phi_m$, allowing the shared curve to follow broad epidemic waves \citep{eilers1996flexible}.

The site-level random intercepts $u_{ij}$ capture persistent differences that are not explained by the shared curve or regional process. Wastewater sites can differ in catchment population, sewer infrastructure, flow conditions, sampling protocol, and laboratory processing. Each $u_{ij}$ shifts its site's overall level across the study period.

The residual process $e_{ij}(t)$ captures within-site temporal dependence and measurement variation after accounting for the other components. Its variance is $\sigma_e^2$, while $\delta_e>0$ controls how slowly correlation decays with temporal separation.

For the sampling times of site $(i,j)$, the Gaussian process specification implies
\[
\bm e_{ij}\sim \MVN(\bm 0,\sigma_e^2R_{ij}),
\qquad
(R_{ij})_{ab}=\exp\{-|t_{ija}-t_{ijb}|/\delta_e\}.
\]
Here $R_{ij}$ is the $T_{ij}\times T_{ij}$ within-site correlation matrix, and $a,b=1,\ldots,T_{ij}$ index two observed times at that site.
Because $R_{ij}$ is constructed from the actual sampling times for each site, the likelihood can accommodate ragged observation windows and irregular sampling without requiring a balanced site-by-time data array.

\textbf{Remarks on potential covariates.}
The formulation above does not include covariates explicitly, but covariate information can be incorporated naturally when reliable covariates are available. For example, time-varying covariates could be included in the shared temporal component $g(t)$ to explain part of the overall wastewater trend, replacing a purely time-indexed mean curve with a regression structure. Similarly, site-level covariates such as catchment size, wastewater flow characteristics, sampling protocol, or laboratory information could be included in the site-level component so that part of the persistent site-to-site heterogeneity is explained directly rather than absorbed by $u_{ij}$.

\subsection{Spatial formulation for regional deviations}
\label{sec:spatial-formulations}

At each modeled time $t$, the vector
$\bm v(t)=(v_1(t),\ldots,v_K(t))^{\top}$ contains the simultaneous regional
departures from the shared curve. We assume
$\bm v(t)\sim\MVN\{\bm0,\Sigma_v\}$ and specify $\Sigma_v$ directly. In the Ontario
system, the analysis is organized by public health units, multiple
sites may contribute to one unit, and exact catchment boundaries are not
consistently available. This motivates an areal covariance based on regional
adjacency.

Let $W=(w_{ii'})$ be a symmetric adjacency matrix, with $w_{ii'}=1$ when regions $i$ and $i'$ are neighbors and $w_{ii}=0$. Write $d_i=\sum_{i'}w_{ii'}$ and $D_W=\operatorname{diag}(d_1,\ldots,d_K)$. For regions with positive degree, define
\[
W^*=D_W^{-1/2}WD_W^{-1/2}.
\]
As in longitudinal marginal models, we specify marginal variance and correlation separately \citep{liang1986longitudinal}. Here, this principle is applied to regional deviations using adjacency to define spatial association. The first-order marginal covariance describes local contemporaneous dependence in regional departures from the shared temporal trend, while keeping regional variability and neighboring association separately interpretable. The same hierarchical framework accommodates other valid spatial covariance structures suited to the surveillance setting. The first-order marginal autoregressive (MAR) model is
\begin{equation}
\Sigma_v=\sigma_v^2(I+\rho W^*),\qquad -1<\rho<1.
\label{eq:mar-spatial}
\end{equation}
The eigenvalues of $W^*$ lie in $[-1,1]$, so the covariance is positive definite for $|\rho|<1$. Negative and positive values allow negative and positive marginal association between adjacent regions, respectively. For an isolated region, its row and column in $W^*$ can be set to zero. The resulting marginal variances and covariances are
\[
\operatorname{Var}\{v_i(t)\}=\sigma_v^2,\qquad
\operatorname{Cov}\{v_i(t),v_{i'}(t)\}
=\frac{\sigma_v^2\rho\,w_{ii'}}{\sqrt{d_id_{i'}}},\quad i\ne i'.
\]
Thus $\sigma_v$ is a common marginal regional standard deviation, while $\rho$ controls the strength of contemporaneous association between neighboring regional departures. The coefficient $\rho$ is not itself the correlation of every neighboring pair. That correlation is $\rho/\sqrt{d_id_{i'}}$. Non-neighboring regional deviations are marginally independent at the same time under this model.

The regional vectors $\bm v(t)$ are independent across modeled time points, conditional on the covariance parameters. The site effects, regional vectors, and site-specific residual processes are mutually independent, and residual processes are independent across sites.
\begin{samepage}
Conditional on the spline coefficients $\bm\phi$, which determine the shared mean curve, the covariance of two responses is
\begin{eqnarray}
    && \operatorname{Cov}\{Y_{ij}(t),Y_{i'j'}(s)\mid\bm\phi\}\nonumber\\
&=& \mathbf 1(t=s)(\Sigma_v)_{ii'}+\mathbf 1(i=i',j=j')
\{\sigma_u^2+\sigma_e^2\exp(-|t-s|/\delta_e)\}.
\label{eq:response-pairwise-cov}
\end{eqnarray}
\end{samepage}

Wastewater surveillance commonly includes multiple sites within one reporting region \citep{li2023wastewater}. Different sites within one region share a regional deviation when observed at the same time. Neighboring regions share local spatial variation. Repeated measurements from a site share both its persistent intercept and a decaying residual process. These components describe distinct covariance patterns. Site-specific residuals cannot represent shared fluctuations across sites, and persistent site intercepts cannot represent changing regional departures. Retaining the hierarchy therefore allows multiple sites to inform a regional trajectory while preserving their own levels and temporal dependence.

\subsection{Regional time-shift extension}
\label{sec:model-delay}

Model~1 describes departures from a shared temporal trajectory. Model~2 additionally allows that trajectory to occur earlier or later in each region within a selected epidemic period.
\[
Y_{ij}(t)=g(t-\delta_i)+v_i(t)+u_{ij}+e_{ij}(t).
\]
All covariance components retain their definitions from Model~1. Setting every $\delta_i$ to zero recovers Model~1.

The shared curve remains a cubic B-spline,
$g(x)=\sum_{m=1}^M B_m(x)\phi_m$, evaluated at $t-\delta_i$. A positive $\delta_i$ shifts the regional mean curve to a later time relative to the common curve.

When $t-\delta_i$ falls outside the study interval, $g(t-\delta_i)$ is evaluated by extending the cubic polynomial on the corresponding boundary segment.

We center shifts across regions, so that no PHU is assigned a fixed reference shift. Let $\bm1$ be the $K$-vector of ones and $H$ a $K\times(K-1)$ matrix with orthonormal columns orthogonal to $\bm1$. We use the standard orthonormal sum-to-zero construction \citep{stan2026reference}. The timing layer is
\[
\bm\delta=\sigma_\delta H\bm z,\qquad
\bm z\sim N(\bm0,I_{K-1}),\qquad \sum_{i=1}^K\delta_i=0.
\]
This construction treats all regions symmetrically and gives a zero-sum Gaussian distribution with
\[
\operatorname{Cov}(\bm\delta\mid\sigma_\delta)
=\sigma_\delta^2HH^\top.
\]
The centering constraint anchors the common curve to the average regional timing. The parameter $\sigma_\delta$ controls regional timing variability. Positive and negative shifts indicate later and earlier alignment with the common curve, respectively.Translating a common curve allows regional leads and lags to be estimated in weeks without attributing them to site levels or residual fluctuations. This shift model draws on Bayesian curve registration \citep{telesca2008registration}.

\subsection{Estimation procedure}
\label{sec:estimation}

For site $(i,j)$, let $T_{ij}$ be observed measurements in $\bm Y_{ij}=(Y_{ij}(t_{ij1}),\ldots,Y_{ij}(t_{ijT_{ij}}))^{\top}$. Let $\bm v_i(\mathcal T_{ij})$ and $\bm g_{ij}(\bm\delta)$ contain the regional effects and shared-curve values at these same times. The latter has entries $g(t_{ija})$ under Model~1 and $g(t_{ija}-\delta_i)$ under Model~2. Given the regional and site effects, the site observation vectors are independent, with
\[
\bm Y_{ij}\mid\bm v_i(\mathcal T_{ij}),u_{ij},\bm\phi,\delta_i,\sigma_e,\delta_e
\sim\MVN\left\{\bm g_{ij}(\bm\delta)+\bm v_i(\mathcal T_{ij})+u_{ij}\bm 1_{T_{ij}},\ \sigma_e^2R_{ij}\right\}.
\]
Here $\bm 1_{T_{ij}}$ is a vector of ones, so the site intercept applies at every observed time. The regional vectors are independent across modeled times given $\sigma_v$ and $\rho$, and the site intercepts are independent given $\sigma_u$.

Stacking the site vectors into $\bm Y\in\mathbb R^N$ and integrating out $\{\bm v(t)\}$ and $\bm u$ gives a multivariate normal distribution with mean formed by the corresponding $\bm g_{ij}(\bm\delta)$ and covariance $\Sigma_Y$. Its entries are given by \eqref{eq:response-pairwise-cov}, with $\Sigma_v$ specified by \eqref{eq:mar-spatial}. We retain the latent variables in the joint posterior to estimate regional and site effects along with the model parameters.

The working priors are
\[
\phi_m\overset{\mathrm{ind}}{\sim} N(0,1.5^2),
\qquad
\sigma_v,\sigma_u,\sigma_e\sim t_3^+(0,2.5),
\]
\[
\rho\sim \mathrm{Uniform}(0,0.99),
\qquad
\delta_e\sim \mathrm{Lognormal}\{\log(10),1\},
\qquad \sigma_\delta\sim N^+(0,3^2)\quad\text{for Model~2}.
\]
Here $t_3^+(0,2.5)$ denotes a half Student-$t$ distribution with 3 degrees of freedom and scale 2.5, and $N^+(0,3^2)$ is a half-normal distribution.
These proper, weakly regularizing priors are specified on the scale of the log wastewater measurements and the time units of each analysis. We use them in the reported real-data and simulation fits; the covariance framework itself does not require these particular prior choices.

Let $\Theta$ collect the covariance parameters, spline coefficients $\bm\phi$, regional vectors $\{\bm v(t)\}$, site intercepts $\bm u$, and, under Model~2, the standardized timing coordinates $\bm z$ and their scale $\sigma_\delta$, which determine $\bm\delta=\sigma_\delta H\bm z$. For Model~2, the joint posterior factorizes as
\[
\begin{aligned}
\pi(\Theta\mid\bm Y)\propto{}&
\left[\prod_{i=1}^K\prod_{j=1}^{n_i}
p\{\bm Y_{ij}\mid\bm v_i(\mathcal T_{ij}),u_{ij},\bm\phi,\delta_i,\sigma_e,\delta_e\}\right]
\left[\prod_{t\in\mathcal T_{\mathrm{full}}}\pi\{\bm v(t)\mid\sigma_v,\rho\}\right]\\
&\times\pi(\bm u\mid\sigma_u)
\pi(\bm z)
\pi(\bm\phi)\pi(\sigma_v)\pi(\rho)\pi(\sigma_u)\pi(\sigma_e)\pi(\delta_e)\pi(\sigma_\delta).
\end{aligned}
\]
For Model~1, set $\delta_i=0$ and omit $\bm z$ and $\sigma_\delta$ and their priors. We sample the joint posterior in Stan using Hamiltonian Monte Carlo with the No-U-Turn sampler. For each of the four period-specific Ontario fits, we run four chains with 2,000 iterations per chain, consisting of 1,000 warmup (burn-in) iterations followed by 1,000 retained iterations, yielding 4,000 posterior draws. Each $R_{ij}$ uses that site's actual observation times, so balanced sampling is not required. The general marginal covariance permits $-1<\rho<1$. The reported fits use the nonnegative working prior above, $\rho\sim\mathrm{Uniform}(0,0.99)$, to represent positive regional association; this inferential restriction is distinct from the sufficient positive-definiteness condition $|\rho|<1$.

\section{Ontario Wastewater Case Study}
\label{sec:case-study}

We fit Model~1 and Model~2 to the same site-week observations within each of two Omicron periods. Comparing the resulting regional shifts and variance components shows what the timing layer adds to the Ontario analysis. A baseline-model analysis of the complete surveillance period is reported in Web Appendix B.

\subsection{Regional timing in two Omicron periods}
\label{sec:case-delay}

A single shift over the complete surveillance period would combine potentially different regional timing patterns across epidemic waves. We therefore analyze two periods separately. The first covers the first year of Omicron from November 30, 2021 to January 2, 2023. The second covers the XBB-and-later period from January 3, 2023 to July 29, 2024.

To estimate shifts from adequately observed trajectories, we require each site to contribute at least 10 observed weeks in a period and each PHU to contribute at least 20 observed region-weeks with 60\% period coverage. We retain PHUs eligible in both periods so the spatial comparison uses the same geography. These criteria reduce the full-record set of 33 PHUs and 56 sites to 30 PHUs and 50 sites in the period analyses. The first period contains 2,779 observations over 57 modeled weeks and uses eight spline basis functions. The second contains 4,070 observations over 82 weeks and uses ten. These basis dimensions were selected through preliminary analyses, which indicated little change in the fitted curves once enough basis functions were included to capture the broad epidemic waves.

Table~\ref{tab:period-scalars} reports posterior means and 90\% credible intervals for the regional spatial parameters, site and residual standard deviations, temporal correlation range, and timing scale in both periods. Under Model~2, the posterior mean regional scales $\sigma_v$ are 0.256 and 0.240, and the one-week residual correlations $\exp(-1/\delta_e)$ are 0.552 and 0.450. The spatial coefficients $\rho$ are 0.838 and 0.471, with considerable uncertainty in the second period. These spatial summaries change little between Model~1 and Model~2. Site-level differences and within-site temporal correlation remain substantial in both periods.

The regional deviation scales are smaller than the site and residual scales. Thus, the relatively large first-period $\rho$ describes association among modest regional departures from the common curve, rather than the overall amount of variation in the surveillance signal. Persistent site differences and regional fluctuations have distinct interpretations in comparisons across PHUs.

The full-record baseline analysis uses all 33 PHUs and 56 sites. Its posterior mean spatial coefficient is 0.780, with posterior SD 0.129, and the average neighboring covariance is 0.0190. Web Figure 3 and Web Table 1 report the implied spatial correlation matrix and the posterior parameter summaries.

The posterior mean timing scale $\sigma_\delta$ is 1.188 weeks in the first period, with a 90\% credible interval of $(0.802,1.643)$, compared with 0.347 weeks and $(0.028,0.799)$ in the second. In the first period, allowing shifts also changes the posterior mean residual standard deviation $\sigma_e$ from 1.015 to 0.977 and the residual correlation range $\delta_e$ from 1.845 to 1.687 weeks. The corresponding estimates in the later period are nearly unchanged. The larger first-period timing scale and the regional contrasts below point to more pronounced differences in when the shared curve rises and falls.

\insertplaceholder{Table}{tab:period-scalars}

\begin{figure}[H]
\centering
\includegraphics[width=\textwidth,height=.72\textheight,keepaspectratio]{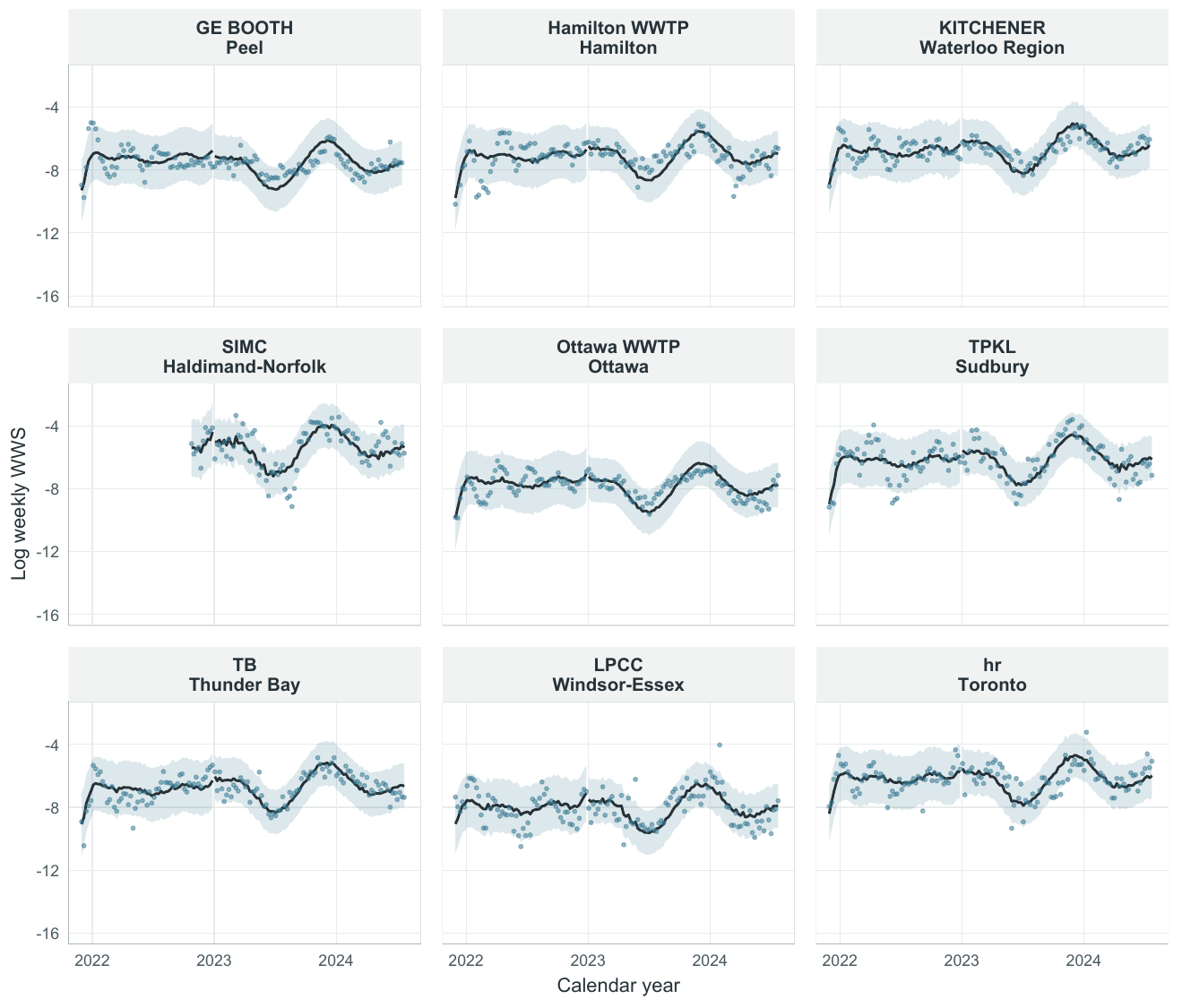}
\caption{Observed weekly log wastewater signals (points), Model~2 posterior mean site-specific latent trajectories (lines), and pointwise 90\% posterior predictive intervals (bands) for the nine sites in Figure~\ref{fig:data-site-series}. }
\label{fig:fitted-site-trajectories}
\end{figure}

Figure~\ref{fig:fitted-site-trajectories} compares the observed signals with Model~2 fitted trajectories for the same nine sites shown in Figure~\ref{fig:data-site-series}. The line represents the posterior mean of the site-specific latent mean, while the band is a pointwise 90\% posterior predictive interval for a replicated weekly observation. Web Figures 8--11 extend the display to all 50 fitted sites. The predictive bands reflect residual variation beyond uncertainty in the latent mean.

\insertplaceholder{Figure}{fig:fitted-site-trajectories}

\begin{figure}[H]
\centering
\includegraphics[width=\textwidth]{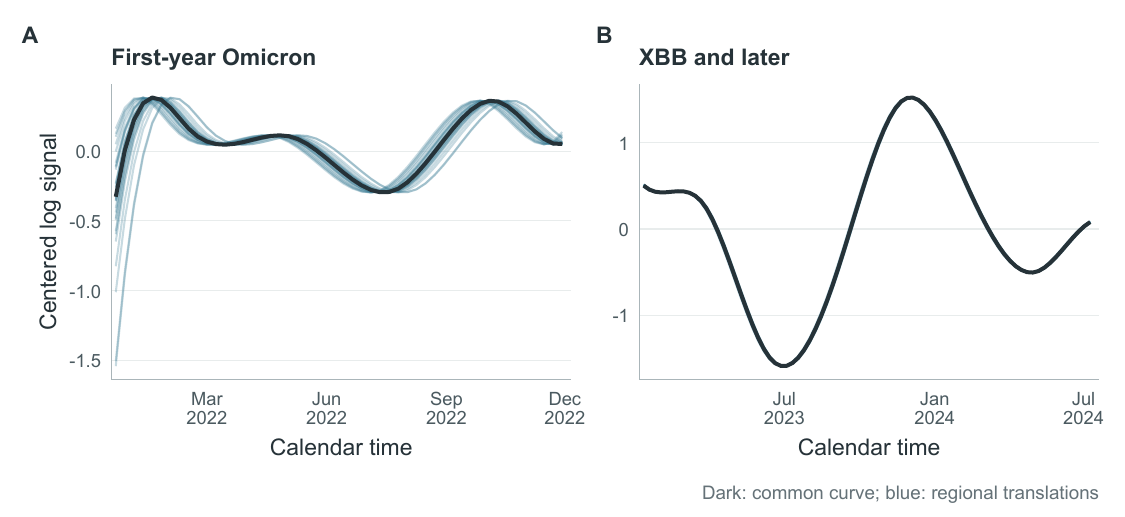}
\caption{Illustration of the estimated time-shift mean curve in the two Ontario periods. The black line is the common curve under the zero-sum shift constraint. Blue lines are translations by the regional posterior mean shifts.}
\label{fig:registered-curves}
\end{figure}

Figure~\ref{fig:registered-curves} translates the estimated shared curve by the regional posterior mean shifts. The first-period regional means separate on the rising and falling portions of the trajectory, while the second-period curves largely overlap. These curves illustrate the time-shift component alone; they exclude $v_i(t)$ and site effects and are not complete fitted wastewater trajectories.

\insertplaceholder{Figure}{fig:registered-curves}

\begin{figure}[H]
\centering
\includegraphics[width=\textwidth,height=.76\textheight,keepaspectratio]{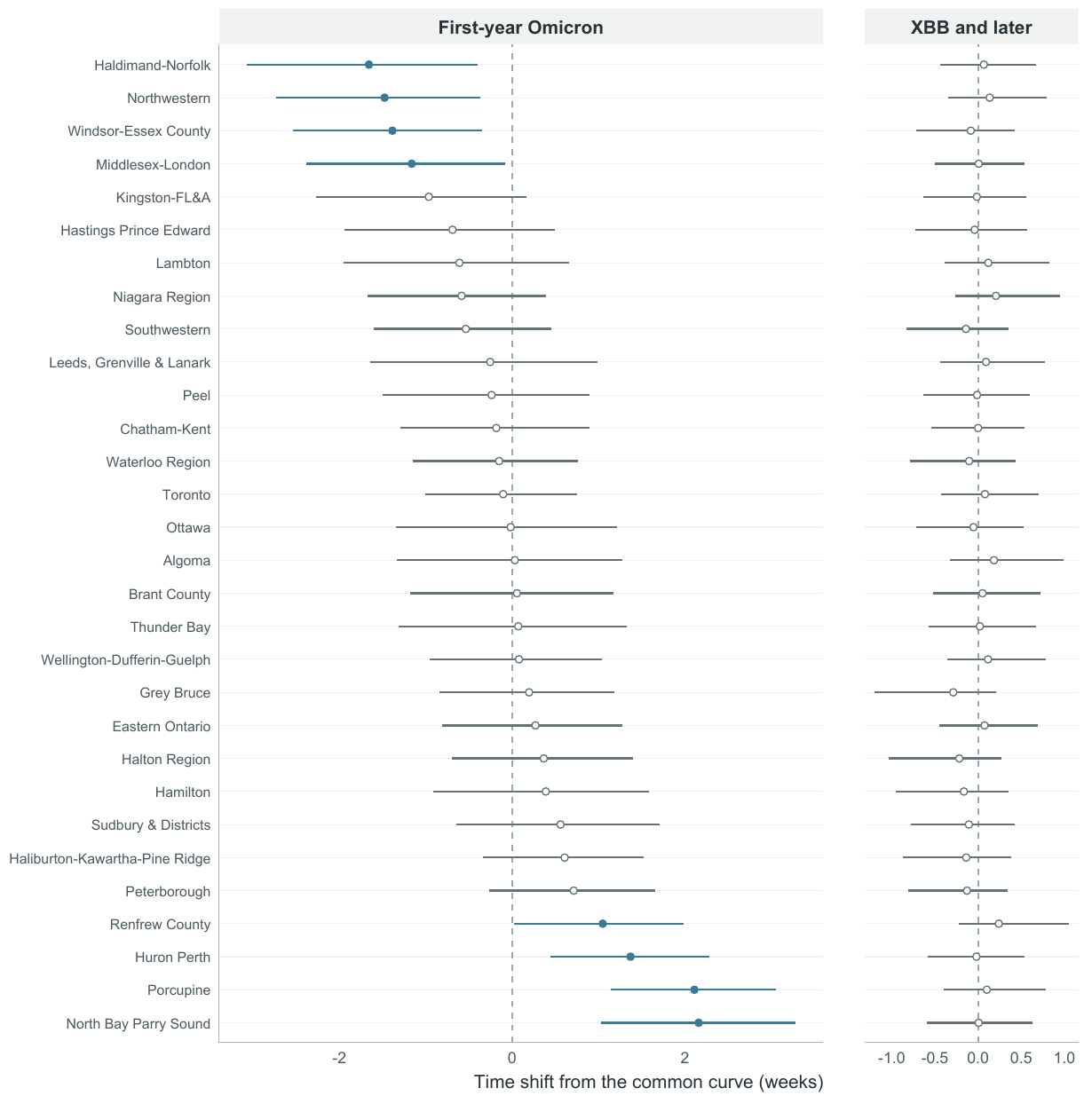}
\caption{Regional posterior mean time shifts and marginal 90\% credible intervals relative to the common curve under the zero-sum shift constraint. Regions are ordered by first-period posterior means. Blue points have intervals excluding zero. All regional shifts are estimated.}
\label{fig:delay-forest}
\end{figure}

The posterior means and 90\% intervals for $\delta_i$ are shown in Figure~\ref{fig:delay-forest}. In the first period, four PHUs have intervals entirely below zero. These are Haldimand-Norfolk, Northwestern, Windsor-Essex, and Middlesex-London. Their posterior mean shifts are approximately 1.16 to 1.66 weeks earlier than the common curve. Four PHUs have intervals entirely above zero. These are Renfrew County and District, Huron Perth, Porcupine, and North Bay Parry Sound, with posterior mean shifts of approximately 1.05 to 2.16 weeks later. All regional intervals include zero in the second period, although uncertainty remains about the size of timing differences. These contrasts describe alignment with the common curve under the zero-sum constraint, not differences from a fixed reference PHU.

Earlier and later regions are interspersed geographically, with no clear province-wide progression. Regional timing was more heterogeneous in the first period than in the second.

\insertplaceholder{Figure}{fig:delay-forest}

\section{Simulation Validation}
\label{sec:simulation}

The simulations assess recovery of the parameters under the observed regional and site structure and quantify how region-specific shifts affect trajectory estimation. The information-size experiment in Web Appendix A extends this assessment across sampling designs.

\subsection{Recovery under the Ontario sampling structure}

The first experiment evaluates frequentist operating characteristics of Bayesian inference under an Ontario-like sampling design. Across repeated datasets, we assess whether posterior means track true parameter values and whether posterior standard deviations reflect the empirical variation of those estimates. We use the observed 33-PHU adjacency graph and the allocation of 56 sites across PHUs. A fixed nonlinear shared mean is represented in fitting by ten cubic B-spline functions. Data are generated from Model~1 with the marginal covariance in \eqref{eq:mar-spatial} and fitted using the working priors in Section~\ref{sec:estimation}. Selected observations are omitted to create irregular within-site schedules. We generate 200 datasets with true parameter values $\rho=0.8$, $\sigma_v=0.5$, $\sigma_u=1.5$, $\sigma_e=1$, and $\delta_e=20$ in units of the simulation time grid. This design preserves the Ontario graph and site allocation but does not reproduce the exact weekly observation schedule.

Table~\ref{tab:application-baseline-simulation} shows close recovery of the regional, site, and residual parameters. The empirical variation of the posterior means is generally similar to the average posterior standard deviation. For the spatial component, the average posterior means of $\rho$ and $\sigma_v$ are 0.787 and 0.502, against true values of 0.8 and 0.5.

Complementary lattice experiments vary the number of regions, sites, and observation times separately (Web Appendix A and Web Figures 1--2). Increasing the repeated time points from 20 to 100 moves the average spatial dependence estimate toward its true value under the MAR specification. This illustrates how temporal replication contributes to learning regional dependence.

\insertplaceholder{Table}{tab:application-baseline-simulation}

\subsection{Validation of the regional time-shift extension}

The second experiment asks whether the regional shifts and other parameters can be recovered when timing differences are present, and what is attributed to the residual process when shifts are ignored. We generate data on the same 33-PHU graph and 56-site allocation with 100 complete weekly observations per site. The shared mean is a smooth two-wave curve projected onto the same ten-dimensional cubic B-spline basis used for fitting. We consider two generating mechanisms. In the first, $\delta_1=0$ and the other shifts are independent $N(0,2^2)$ draws. In the second, all shifts are zero. We generate 200 datasets per mechanism and fit both Model~1 and Model~2, using the same shift parameterization as the generating mechanism, to each dataset, for 800 fits. Validation of the centered timing layer under the observed period-specific sampling pattern follows below. The remaining true values are $\rho=0.8$, $\sigma_v=0.5$, $\sigma_u=1.5$, $\sigma_e=1$, and $\delta_e=20$.

\begin{figure}[H]
\centering
\includegraphics[width=.94\textwidth]{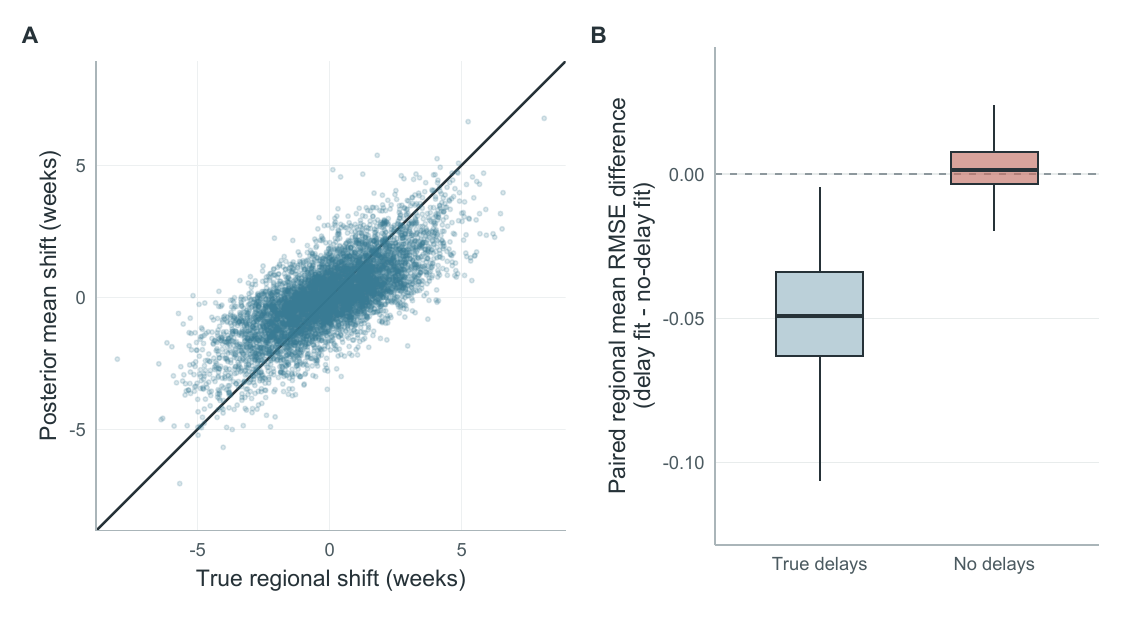}
\caption{Timing simulation with 200 datasets per generating mechanism under the reference-region shift parameterization. (A) True shifts versus their posterior means for PHUs; the line shows equality. (B) Paired regional mean-curve RMSE difference, Model~2 minus Model~1, under true delays (left box) and no delays (right box). Negative values favor Model~2.}
\label{fig:application-delay-simulation}
\end{figure}

When shifts are present, the average posterior mean of $\sigma_\delta$ is 1.885 against a true value of 2, with empirical SD 0.476 and average posterior SD 0.466. The empirical coverage of marginal 90\% intervals for the shifts is 87.0\%. Figure~\ref{fig:application-delay-simulation}A shows recovery of their relative ordering, although the largest shifts are pulled toward the reference.

When the data contain shifts, omitting them increases the average estimates of $\sigma_e$ from 1.011 to 1.046 and $\delta_e$ from 20.63 to 21.85. The residual temporal process then absorbs some unmodeled timing variation, while $\rho$ and $\sigma_v$ change little. Model~2 also improves regional mean-curve recovery, with average root mean squared
error (RMSE) 0.242 rather than 0.292. When the data contain no shifts, the two models recover regional mean curves similarly. Figure~\ref{fig:application-delay-simulation}B shows the paired RMSE difference, Model~2 minus Model~1, across 200 datasets under each generating mechanism. The differences are predominantly negative when shifts are present and close to zero when they are absent. Under Model~2, the average posterior mean of $\sigma_\delta$ is still 0.572 because a continuous prior on a nonnegative scale does not put mass at zero. A positive estimate alone therefore does not establish that shifts are present. The posterior scale and the region-specific contrasts should be interpreted together.

\insertplaceholder{Figure}{fig:application-delay-simulation}

Further validation uses the first-period graph and observed weekly sampling pattern, with 200 datasets each under no shifts, true shifts, and wave-height differences without shifts. With regional shifts centered across PHUs, Model~2 estimates the true timing scale of 2 weeks at 2.005 weeks on average. The 90\% posterior credible intervals for individual regional shifts achieve 90.5\% empirical coverage. Omitting true shifts increases the estimated time-residual scale and range, whereas wave-height differences alone produce few nonzero timing contrasts in this design. Different intensities of common waves therefore do not affect the estimates of regional leads or lags. The shared-trend formulation is motivated by the broad waves observed across Ontario sites in Figure~\ref{fig:data-site-series}, with regional departures and site effects representing local variation. Web Appendix D and Web Tables 2 and 3 give the design and results.

Together, the simulations support estimation of the baseline decomposition under the Ontario sampling hierarchy and show how unmodeled timing can be absorbed by the residual process.

\section{Discussion}
\label{sec:discussion}

This work develops a Bayesian spatio-temporal framework for multi-site wastewater surveillance. Its principal contribution is a common statistical basis for distinguishing where surveillance signals differ, which sampling level accounts for those differences, and when regional waves occur. Retaining sites within PHUs separates persistent site levels from evolving regional dynamics, making regional comparisons interpretable despite heterogeneous site records. The spatial covariance distinguishes the magnitude of regional departures, through $\sigma_v$, from their neighboring association, through $\rho$. This provides complementary summaries of local fluctuation and regional synchrony. Regional adjacency provides a practical spatial representation when wastewater service areas are incompletely known.

The Ontario analyses show persistent site differences and residual temporal dependence in both Omicron periods. In the first year of Omicron, the larger timing scale and several region-specific contrasts indicate meaningful differences in when the shared trajectory occurred across PHUs. The temporal residual scale and range also decrease when shifts are included, while regional spatial parameters remain similar. In the XBB-and-later period, the timing scale is smaller and individual contrasts are less pronounced. These findings distinguish regional timing from differences in signal level and from within-site temporal fluctuation. They also show that regional timing summaries are specific to the epidemic period, rather than fixed characteristics of a PHU.

The joint separation of shared regional fluctuations, persistent site differences, and within-site temporal dependence remains less explicit in existing wastewater models \citep{li2023wastewater,somerset2025differentiable}. The Ontario-inspired simulations support recovery of these components under the site-within-region sampling hierarchy (Table~\ref{tab:application-baseline-simulation} and Web Appendix~A). This matters when multiple sites inform one public health region, since sustained differences among their measurements should be distinguished from changes shared across that region. The timing experiments further show that omitting true shifts assigns systematic regional misalignment to larger and more persistent within-site fluctuations. When shifts are absent, the models give similar regional trajectories. Together, these results support interpreting regional surveillance through distinct sources of variation rather than through curve fit alone.

Further methodological work can clarify how curve specification and boundary handling affect regional timing summaries. Because the shifts align a common curve over a selected period, their interpretation depends on the fitted functional form and its treatment outside the observation interval. Evaluating alternative smoothing choices, analysis windows, and extrapolation rules would clarify the stability of the resulting shifts under the boundary cubic continuation used here and other boundary treatments.

The framework could also support surveillance where local sampling is limited. Wastewater monitoring in small communities provides one relevant application setting \citep{pappu2025tracking}. The shared curve pools information across the network, while neighboring observations can inform contemporaneous regional deviations. Evaluation under sparse sampling could establish when this information sharing supports useful local summaries and how uncertainty changes with monitoring coverage.

The same hierarchy provides a structure for explaining heterogeneity through covariates at the corresponding sampling level. Catchment population, sewer infrastructure, flow, and laboratory or assay characteristics could help explain persistent site differences, while rainfall and operational changes could enter time-varying measurement components. Regional population density, age composition, mobility, vaccination, and prior-infection indicators could be related to regional dynamics and timing. Incorporating these variables within the hierarchy would carry uncertainty in the latent components into the estimated associations.

Linking wastewater trajectories with regional hospital admissions through a lagged association would connect these surveillance summaries to clinical outcomes. A joint model could allow regional shifts to vary across waves and relate them to the covariates above. Validation on subsequent waves could assess whether the resulting model anticipates local increases in hospital admissions.

\section*{Acknowledgments}
The study was partially supported by the CIHR Applied Public Health Chair in Environment, Climate Change and One Health, awarded to Dr. Robert Delatolla, and by the Canadian Statistical Sciences Institute (CANSSI) Collaborative Research Team (CRT) in ``The Application of Statistical Methods to Wastewater Analysis." The authors acknowledge data shared by Ontario Wastewater Surveillance Consortium.

\section*{Data Availability}
The Ontario wastewater surveillance program and measurement methods are described by \citet{daoust2024viral}. Publicly released data are available in CSV format from Zenodo (\url{https://doi.org/10.5281/zenodo.10810195}) and the Ontario Wastewater Surveillance Consortium GitHub repository (\url{https://github.com/OntarioWastewaterSurveillanceConsortium/sars-cov-2-data}).

\bibliographystyle{plainnat}
\bibliography{Bibliography-MM-MC}

\section*{Supplementary Material}
Web Appendices A--D, Web Tables 1--3, and Web Figures 1--11, referenced in Sections 2, 4, 5, and 6, are provided in the accompanying Supplementary Material.

\section*{Tables}

\begin{table}[H]
\centering
\footnotesize
\setlength{\tabcolsep}{0.8pt}
\caption{Posterior means and 90\% credible intervals from the Ontario period analyses. Both periods use 30 PHUs and 50 sites, with regional shifts centered across PHUs.}
\label{tab:period-scalars}

\begin{tabular}{l r l r l r l r l}
\toprule
& \multicolumn{4}{c}{First-year Omicron}
& \multicolumn{4}{c}{XBB and later} \\
\cmidrule(lr){2-5}\cmidrule(lr){6-9}

& \multicolumn{2}{c}{Model 1}
& \multicolumn{2}{c}{Model 2}
& \multicolumn{2}{c}{Model 1}
& \multicolumn{2}{c}{Model 2} \\
\cmidrule(lr){2-3}\cmidrule(lr){4-5}
\cmidrule(lr){6-7}\cmidrule(lr){8-9}

Parameter
& \multicolumn{1}{c}{Estimate}
& \multicolumn{1}{c}{90\% CrI}
& \multicolumn{1}{c}{Estimate}
& \multicolumn{1}{c}{90\% CrI}
& \multicolumn{1}{c}{Estimate}
& \multicolumn{1}{c}{90\% CrI}
& \multicolumn{1}{c}{Estimate}
& \multicolumn{1}{c}{90\% CrI} \\
\midrule

$\sigma_v$
& 0.249 & (0.198, 0.298)
& 0.256 & (0.202, 0.305)
& 0.240 & (0.202, 0.276)
& 0.240 & (0.202, 0.277) \\

$\sigma_u$
& 1.083 & (0.901, 1.294)
& 1.123 & (0.946, 1.343)
& 0.912 & (0.768, 1.085)
& 0.909 & (0.765, 1.087) \\

$\sigma_e$
& 1.015 & (0.976, 1.054)
& 0.977 & (0.940, 1.016)
& 0.809 & (0.787, 0.832)
& 0.808 & (0.786, 0.831) \\

$\rho$
& 0.807 & (0.474, 0.980)
& 0.838 & (0.549, 0.983)
& 0.460 & (0.085, 0.891)
& 0.471 & (0.078, 0.884) \\

$\delta_e$ (weeks)
& 1.845 & (1.678, 2.031)
& 1.687 & (1.529, 1.855)
& 1.257 & (1.165, 1.352)
& 1.253 & (1.162, 1.353) \\

$\exp(-1/\delta_e)$
& 0.581 & (0.551, 0.611)
& 0.552 & (0.520, 0.583)
& 0.451 & (0.424, 0.477)
& 0.450 & (0.423, 0.478) \\

$\sigma_\delta$ (weeks)
& -- & --
& 1.188 & (0.802, 1.643)
& -- & --
& 0.347 & (0.028, 0.799) \\

\bottomrule
\end{tabular}
\end{table}

\begin{table}[H]
\centering
\small
\caption{Baseline-model recovery under the Ontario-inspired 33-unit, 56-site design using the MAR covariance. Mean is the average posterior mean, EmpSD the empirical standard deviation of posterior means, and AvgPostSD the average posterior standard deviation across 200 datasets.}
\label{tab:application-baseline-simulation}
\setlength{\tabcolsep}{6pt}
\begin{tabular}{lrrrr}
\toprule
Parameter & True & Mean & EmpSD & AvgPostSD\\
\midrule
$\rho$ & 0.800 & 0.787 & 0.104 & 0.091\\
$\sigma_v$ & 0.500 & 0.502 & 0.012 & 0.010\\
$\sigma_u$ & 1.500 & 1.480 & 0.147 & 0.158\\
$\sigma_e$ & 1.000 & 0.996 & 0.040 & 0.042\\
$\delta_e$ & 20.000 & 19.800 & 1.990 & 2.010\\
\bottomrule
\end{tabular}
\end{table}

\section*{Figures}

\end{document}


\maketitle

\section{Baseline simulation design and retained results}
\begin{figure}[H]
\centering
\includegraphics[width=0.55\textwidth]{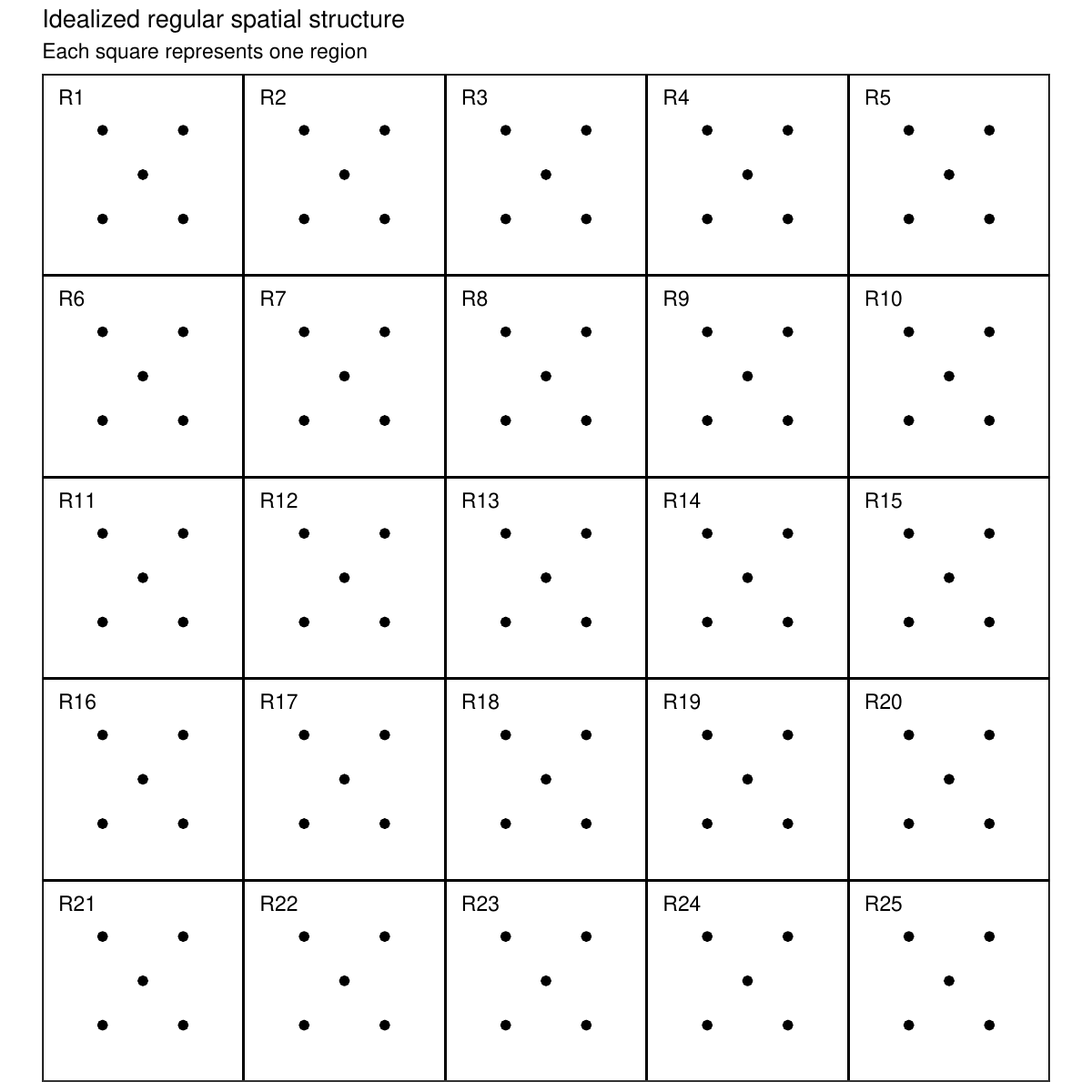}
\caption{Example regular spatial structure used in the idealized simulation settings. The display shows the baseline grid configuration with $K=25$ regions arranged on a $5\times 5$ lattice and $n_i=5$ wastewater sites per region. Lines indicate neighboring regions in the adjacency graph. Other idealized settings change the number of regions, the number of sites per region, or the number of time points, while preserving the same regular grid-based neighborhood structure.}
\label{fig:ideal-spatial-structure}
\end{figure}

\begin{figure}[H]
\centering
\includegraphics[width=\textwidth]{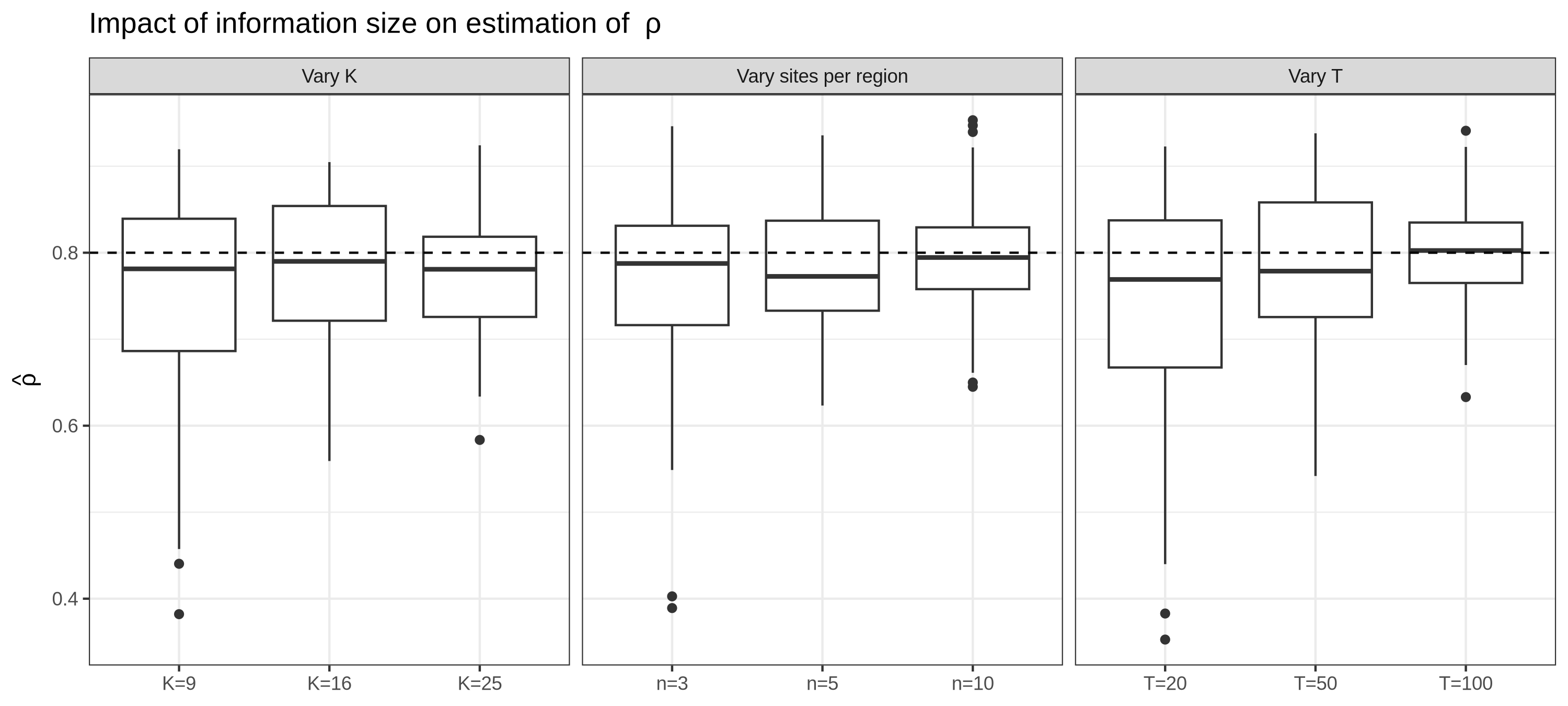}
\caption{Recovery of the MAR spatial dependence parameter in the information-size simulation. The dashed line is the true value $\rho=0.8$. Each panel varies one design feature while holding the others fixed at the baseline values.}
\label{fig:info-size-rho}
\end{figure}

Web Figures~\ref{fig:ideal-spatial-structure} and~\ref{fig:info-size-rho}
summarize the supplementary information-size experiment under the MAR covariance.
Web Figure~\ref{fig:ideal-spatial-structure} shows the baseline $5\times5$ lattice with
five sites per region. Across seven settings, we vary the number of regions,
sites per region, or observation times one at a time. Data are generated and
fitted using the marginal covariance specified in the main text.
Web Figure~\ref{fig:info-size-rho} shows how recovery of the spatial dependence
parameter changes with these information dimensions.

\section{Full-period Ontario analysis}
\begin{figure}[H]
\centering
\includegraphics[width=0.75\textwidth]{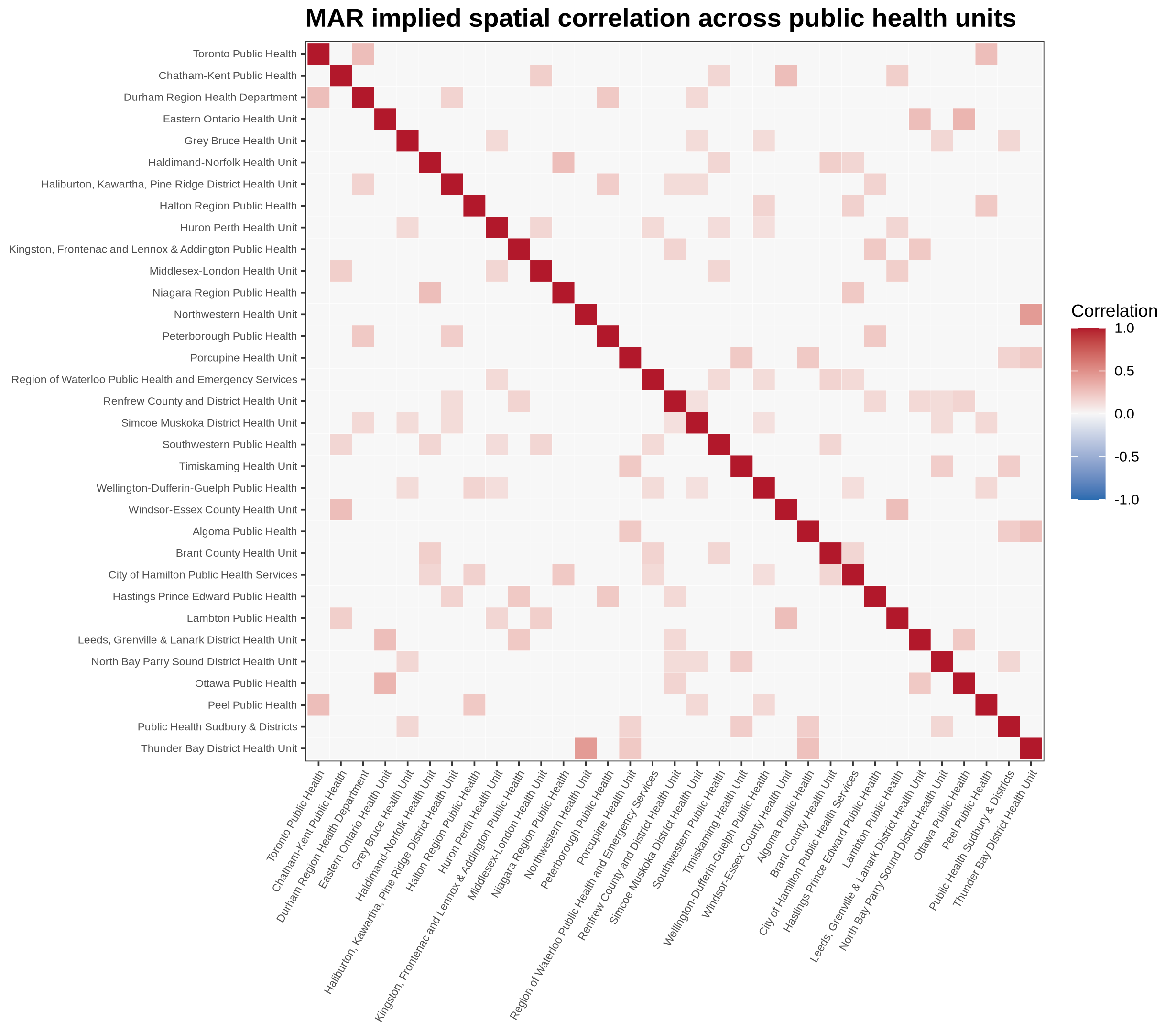}
\caption{Posterior implied regional correlation matrix under the MAR covariance in the full-period 33-unit Ontario analysis. Off-diagonal association is restricted to neighboring public health units by the first-order specification.}
\label{fig:case-study-spatial-correlation}
\end{figure}

The baseline model was also fitted to the full weekly record
using all 33 public health units and 56 wastewater sites.
Web Table~\ref{tab:full-period-posterior} reports posterior summaries under the MAR
covariance. The posterior mean site, regional, and residual standard deviations
were 1.217, 0.319, and 1.252, respectively, and the estimated one-week residual
correlation was 0.709. The posterior mean spatial coefficient was 0.780, with
posterior SD 0.129, and the average neighboring covariance was 0.0190.
Web Figure~\ref{fig:case-study-spatial-correlation} shows the implied spatial
correlation matrix. Non-neighboring regional deviations have zero covariance
under the first-order specification.

\section{Complete site-level trajectories}
\begin{figure}[H]
\centering
\includegraphics[width=\textwidth,height=.77\textheight,keepaspectratio]{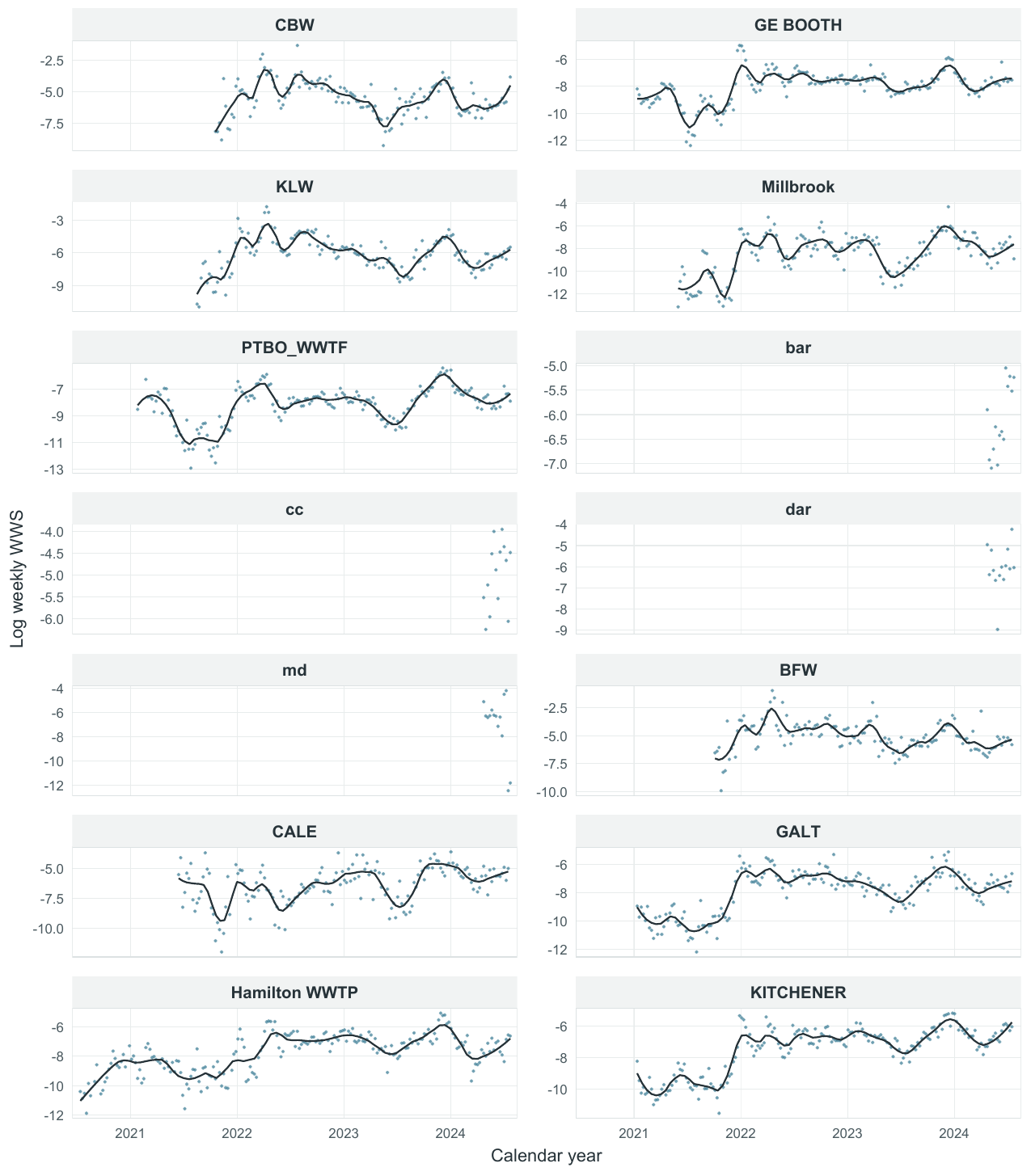}
\caption{Observed weekly log wastewater signals across all sites (part 1 of 4). Lines are descriptive LOESS smooths.}
\label{fig:supp-descriptive-first}
\end{figure}

\begin{figure}[H]
\centering
\includegraphics[width=\textwidth,height=.77\textheight,keepaspectratio]{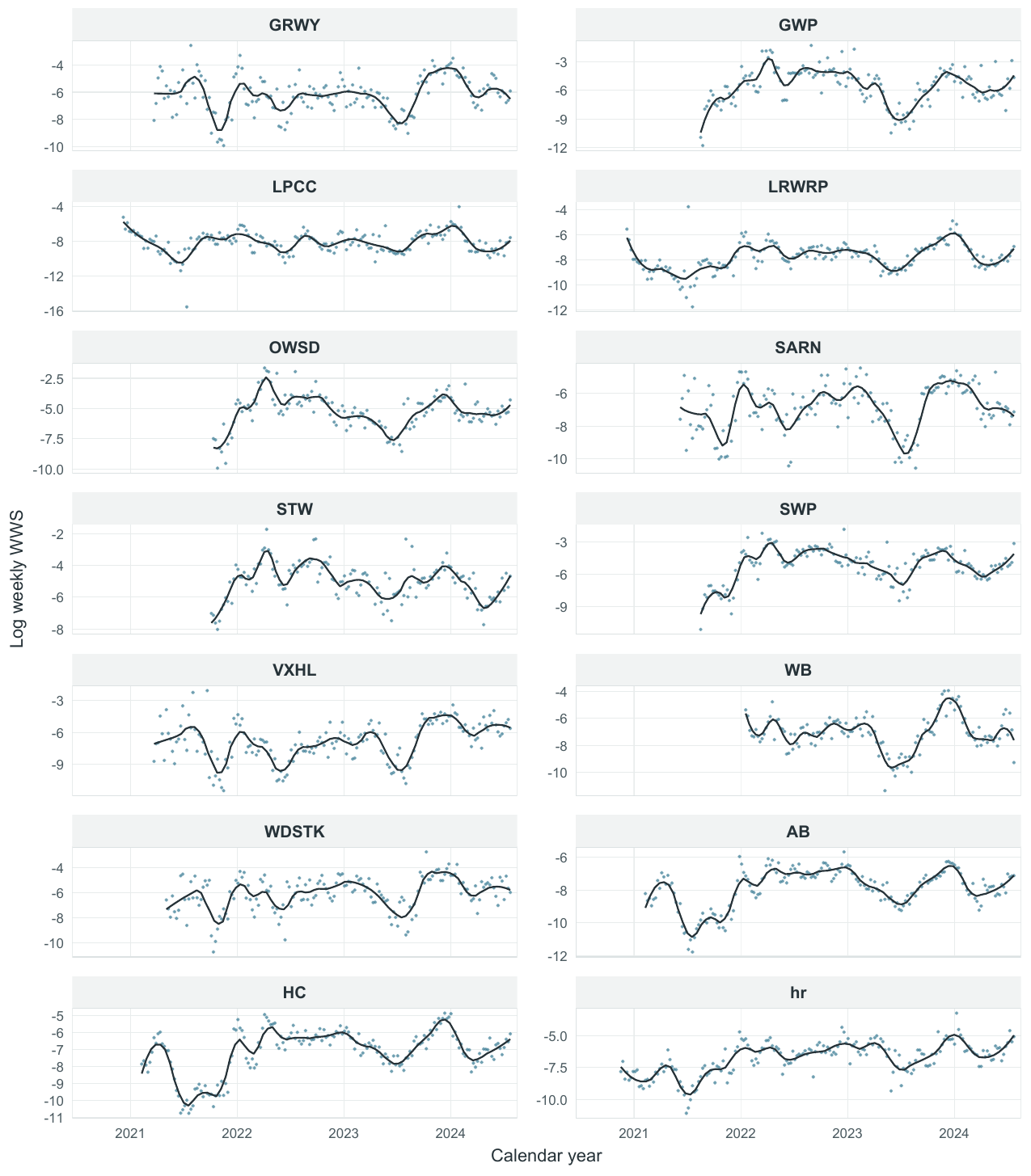}
\caption{Observed weekly log wastewater signals (part 4 of 4).}
\label{fig:supp-descriptive-last}
\end{figure}

\begin{figure}[H]
\centering
\includegraphics[width=\textwidth,height=.77\textheight,keepaspectratio]{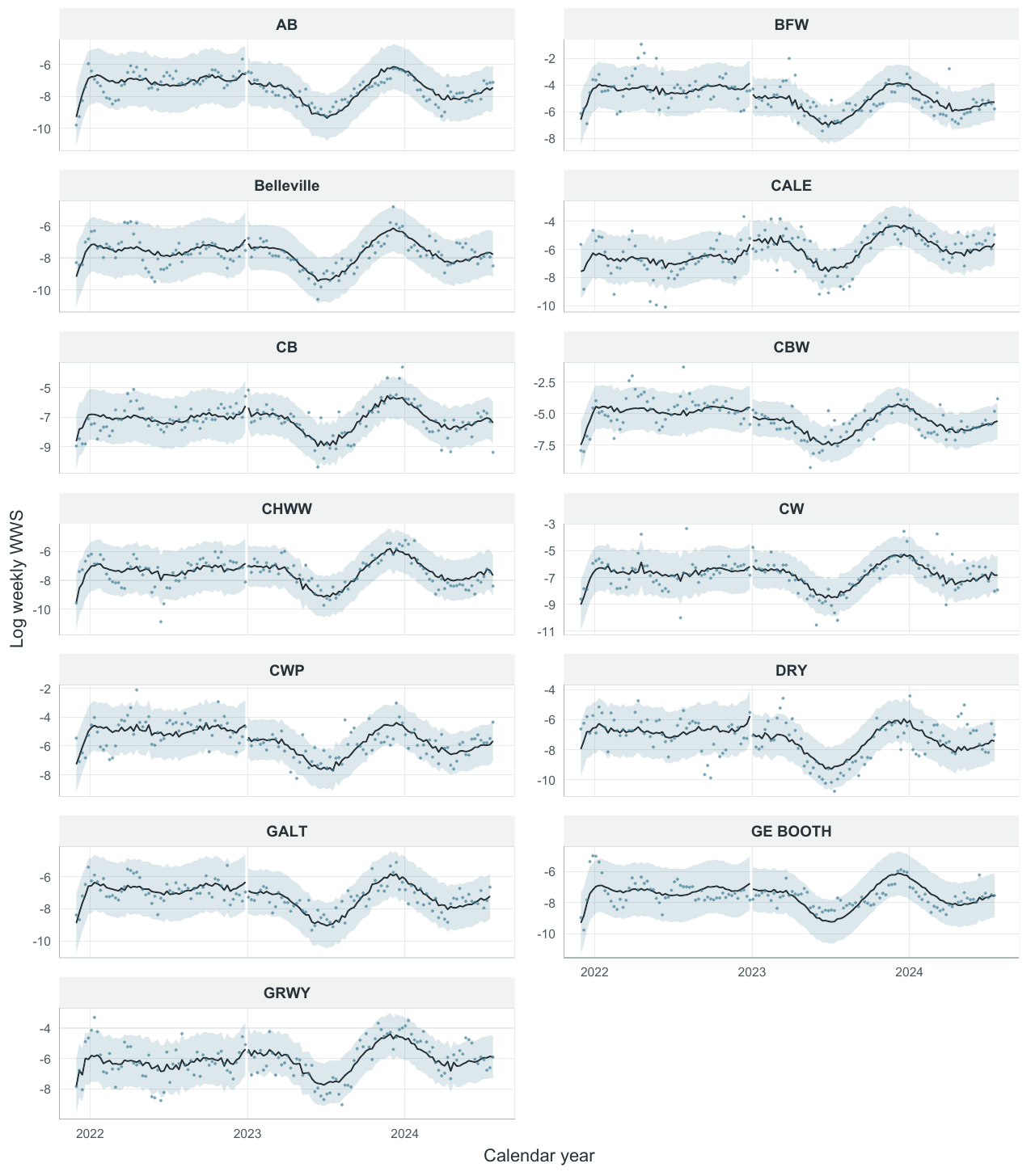}
\caption{Model~2 matched-period fits across all included sites (part 1 of 4). Points are observations, lines are posterior mean latent trajectories, and bands are approximate pointwise 90\% posterior predictive intervals.}
\label{fig:supp-fitted-first}
\end{figure}

\begin{figure}[H]
\centering
\includegraphics[width=\textwidth,height=.77\textheight,keepaspectratio]{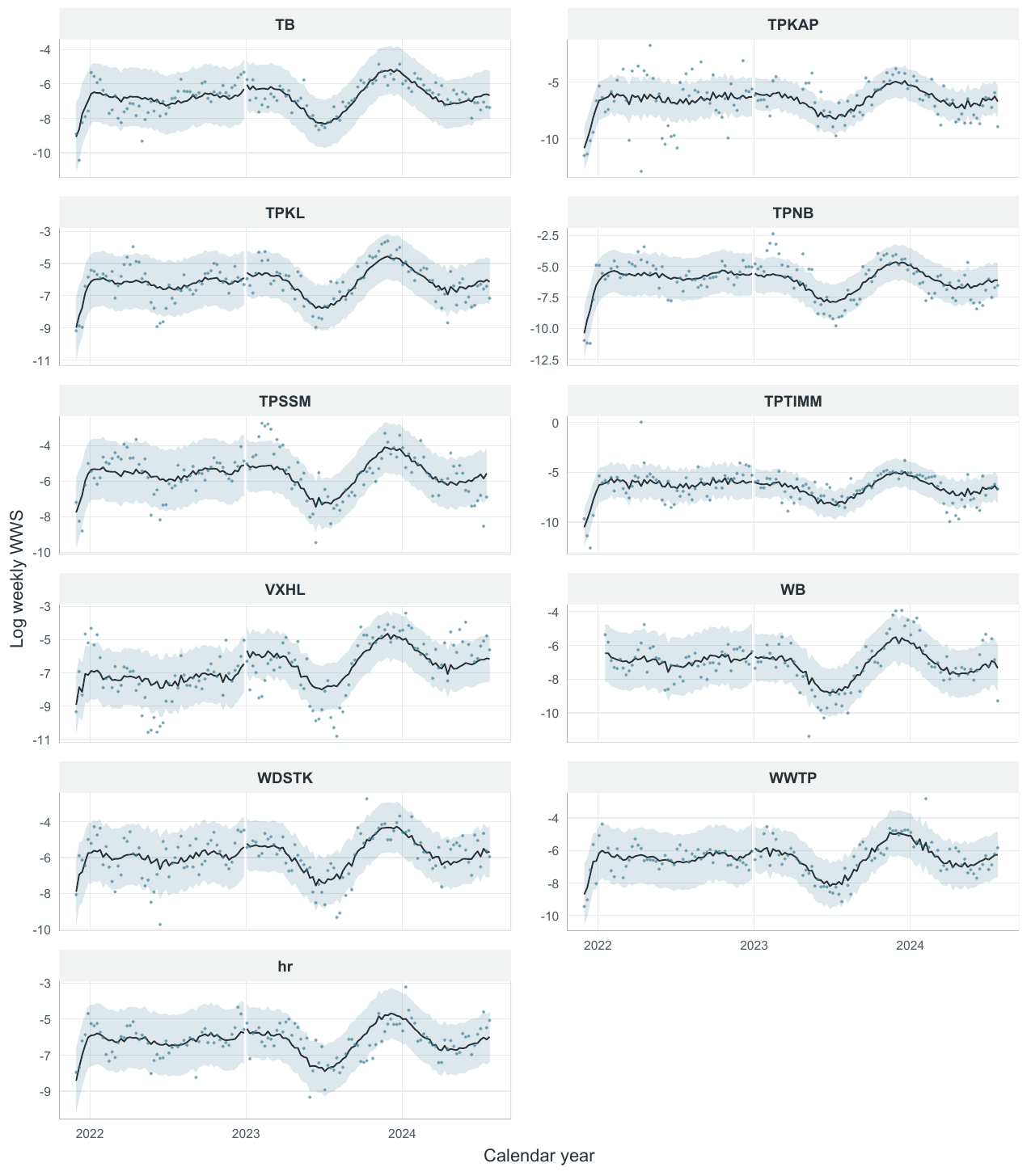}
\caption{Model~2 matched-period site fits (part 4 of 4).}
\label{fig:supp-fitted-last}
\end{figure}

Web Figures~\ref{fig:supp-descriptive-first}--\ref{fig:supp-descriptive-last}
show the descriptive weekly trajectories for all 56 wastewater sites.
Each weekly value is the natural logarithm of the site's
arithmetic mean normalized N1/N2 signal for that week.
Web Figures~\ref{fig:supp-fitted-first}--\ref{fig:supp-fitted-last}
show the fitted trajectories for the 50 sites included in both matched period
analyses. These fits use Model~2 with regional shifts centered
across PHUs. The fitted line is the posterior mean latent site trajectory, while
the band is pointwise 90\% posterior predictive interval for a replicated
weekly observation. Fitted curves are restricted to each site's observed
window within the two analysis periods.

\section{Further validation of regional time shifts}
\label{sec:timing-further-validation}

This simulation study further evaluate recovery of regional time shifts under the Ontario sampling structure and assess whether differences in wave height can be mistaken for timing differences. They also examine how omitting true shifts affects estimation of the temporal residual components.

We generate 200 datasets per mechanism using the first-period design of
30 PHUs, 50 sites, 57 weeks, and 2,779 observed site-week cells, and fit
both models to each dataset. Graph normalization retains degrees from the
full 33-PHU graph. The true values are $\rho=0.8$, $\sigma_v=0.5$,
$\sigma_u=1.5$, $\sigma_e=1$, and $\delta_e=3$ weeks. 

The shared curve uses eight cubic B-spline functions on weeks 1--57, with
internal knots at 12.2, 23.4, 34.6, and 45.8. Define $f_1(t)$ and $f_2(t)$
by centering $1.7B_3(t)$ and $2.1B_6(t)$ over this grid and dividing both
by the sample standard deviation of their uncentered sum. Then
$g(t)=f_1(t)+f_2(t)$ has unit standard deviation, with component peaks at
weeks 13 and 45. The three data generating mechanisms are
\begin{itemize}
\item A has $g_i(t)=g(t)$ and $\delta_i=0$ for every PHU.
\item B has $g_i(t)=g(t-\delta_i)$ with $\bm\delta=2H\bm z$, so the true
      timing scale is $\sigma_\delta=2$ weeks.
\item C has $g_i(t)=(1+a_i)f_1(t)+(1-a_i)f_2(t)$ and no time shifts,
      where $\widetilde a_i\stackrel{\mathrm{iid}}{\sim}U(-0.3,0.3)$ and
      $a_i=\widetilde a_i-K^{-1}\sum_k\widetilde a_k$.
      This varies relative wave heights while preserving their timing.
\end{itemize}
Replications share regional, site, and residual draws across mechanisms. Each fit uses four chains,
each with 1,000 warmup and 500 sampling iterations.

Omitting true shifts inflates the temporal residual scale and range,
while Model~2 recovers both parameters (Web Table~\ref{tab:timing-validation-residual}).
The models give similar estimates when shifts are absent. Across all six
combinations, average estimates are 0.502--0.505 for $\sigma_v$,
1.509--1.524 for $\sigma_u$, and 0.753--0.757 for $\rho$. For $\rho$,
the average posterior SD is approximately 0.119 and 90\% interval coverage
is 95.5--96.5\%.

Model~2 recovers the true timing scale of 2 weeks with an average estimate
of 2.005, individual shift coverage of 90.45\%, and direction accuracy of
95.15\% for shifts of at least one week in absolute value
(Web Table~\ref{tab:timing-validation-shifts}).
Under A and C, only one of 6,000 regional 90\% intervals excludes zero in
each mechanism. These results distinguish genuine time shifts from the
specified wave-height differences.

\clearpage
\section*{Web Tables}

\begin{table}[H]
\centering
\small
\setlength{\tabcolsep}{8pt}
\renewcommand{\arraystretch}{1.12}
\begin{tabular}{lcc}
\toprule
Parameter & Mean & SD \\
\midrule
$\sigma_v$ & 0.319 & 0.015 \\
$\sigma_u$ & 1.217 & 0.234 \\
$\sigma_e$ & 1.252 & 0.018 \\
$\rho$ & 0.780 & 0.129 \\
$\delta_e$ & 2.911 & 0.104 \\
$\exp(-1/\delta_e)$ & 0.709 & 0.009 \\
Fitted-mean RMSE & 1.190 & -- \\
\bottomrule
\end{tabular}
\caption{Full-period posterior summaries and in-sample fitted-mean RMSE under the MAR covariance for the Ontario 33-unit analysis.}
\label{tab:full-period-posterior}
\end{table}

\begin{table}[H]
\centering
\footnotesize
\setlength{\tabcolsep}{5pt}
\caption{Time-residual parameter recovery in the additional validation. The true values are $\sigma_e=1$ and $\delta_e=3$ weeks. A has no shifts, B has true shifts, and C has wave-height differences without shifts. Mean is the average posterior mean, EmpSD is its empirical standard deviation, and AvgPostSD is the average posterior standard deviation across 200 datasets.}
\label{tab:timing-validation-residual}
\begin{tabular}{lcrrrrrr}
\toprule
& & \multicolumn{3}{c}{$\sigma_e$} & \multicolumn{3}{c}{$\delta_e$} \\
\cmidrule(lr){3-5}\cmidrule(lr){6-8}
DGP & Model & Mean & EmpSD & AvgPostSD & Mean & EmpSD & AvgPostSD \\
\midrule
A & 1 & 1.004 & 0.030 & 0.028 & 3.036 & 0.223 & 0.216 \\
A & 2 & 1.001 & 0.030 & 0.029 & 3.014 & 0.225 & 0.216 \\
\addlinespace
B & 1 & 1.101 & 0.046 & 0.035 & 3.719 & 0.348 & 0.285 \\
B & 2 & 1.006 & 0.032 & 0.030 & 3.051 & 0.239 & 0.226 \\
\addlinespace
C & 1 & 1.054 & 0.035 & 0.032 & 3.425 & 0.270 & 0.254 \\
C & 2 & 1.052 & 0.035 & 0.032 & 3.411 & 0.272 & 0.255 \\
\bottomrule
\end{tabular}
\end{table}

\begin{table}[H]
\centering
\footnotesize
\setlength{\tabcolsep}{5pt}
\caption{Time-shift recovery under Model 2 across 200 datasets per DGP. Mean, EmpSD and AvgPostSD summarize $\sigma_\delta$ as in Web Table~\ref{tab:timing-validation-residual}. The 90\% interval coverage is averaged over regions and datasets. Direction accuracy is averaged over datasets for regions with $|\delta_i|\geq1$ week. It is not applicable in A or C, where every true shift is zero.}
\label{tab:timing-validation-shifts}
\begin{tabular}{lrrrrrr}
\toprule
& \multicolumn{4}{c}{$\sigma_\delta$ (weeks)} & \multicolumn{2}{c}{Individual shifts} \\
\cmidrule(lr){2-5}\cmidrule(lr){6-7}
DGP & True & Mean & EmpSD & AvgPostSD & Coverage (\%) & Direction (\%) \\
\midrule
A & 0 & 0.360 & 0.123 & 0.237 & 99.98 & -- \\
B & 2 & 2.005 & 0.341 & 0.353 & 90.45 & 95.15 \\
C & 0 & 0.333 & 0.097 & 0.233 & 99.98 & -- \\
\bottomrule
\end{tabular}
\end{table}

\clearpage
\section*{Web Figures}

\begin{figure}[H]
\centering
\includegraphics[width=\textwidth,height=.77\textheight,keepaspectratio]{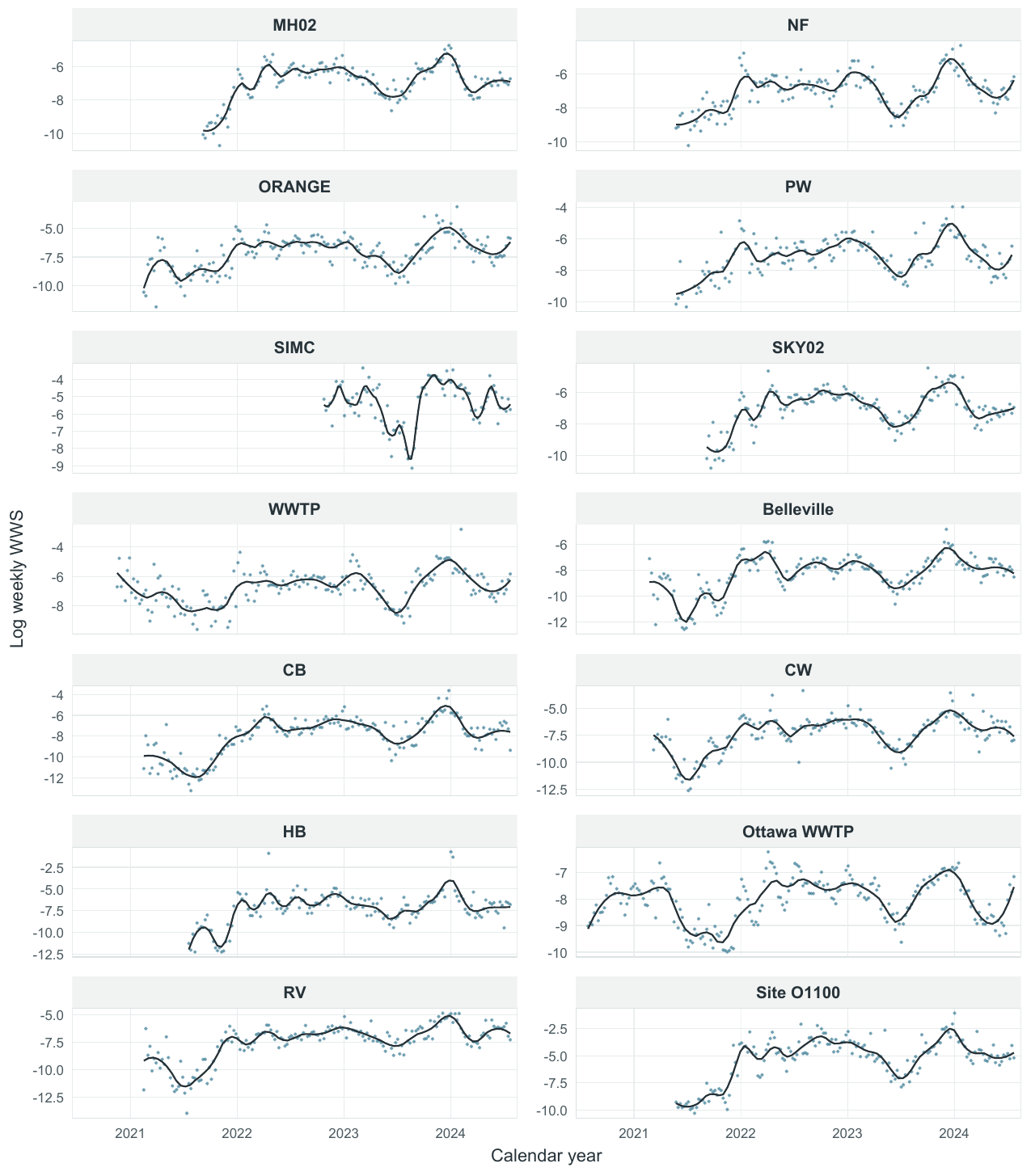}
\caption{Observed weekly log wastewater signals (part 2 of 4).}
\end{figure}

\begin{figure}[H]
\centering
\includegraphics[width=\textwidth,height=.77\textheight,keepaspectratio]{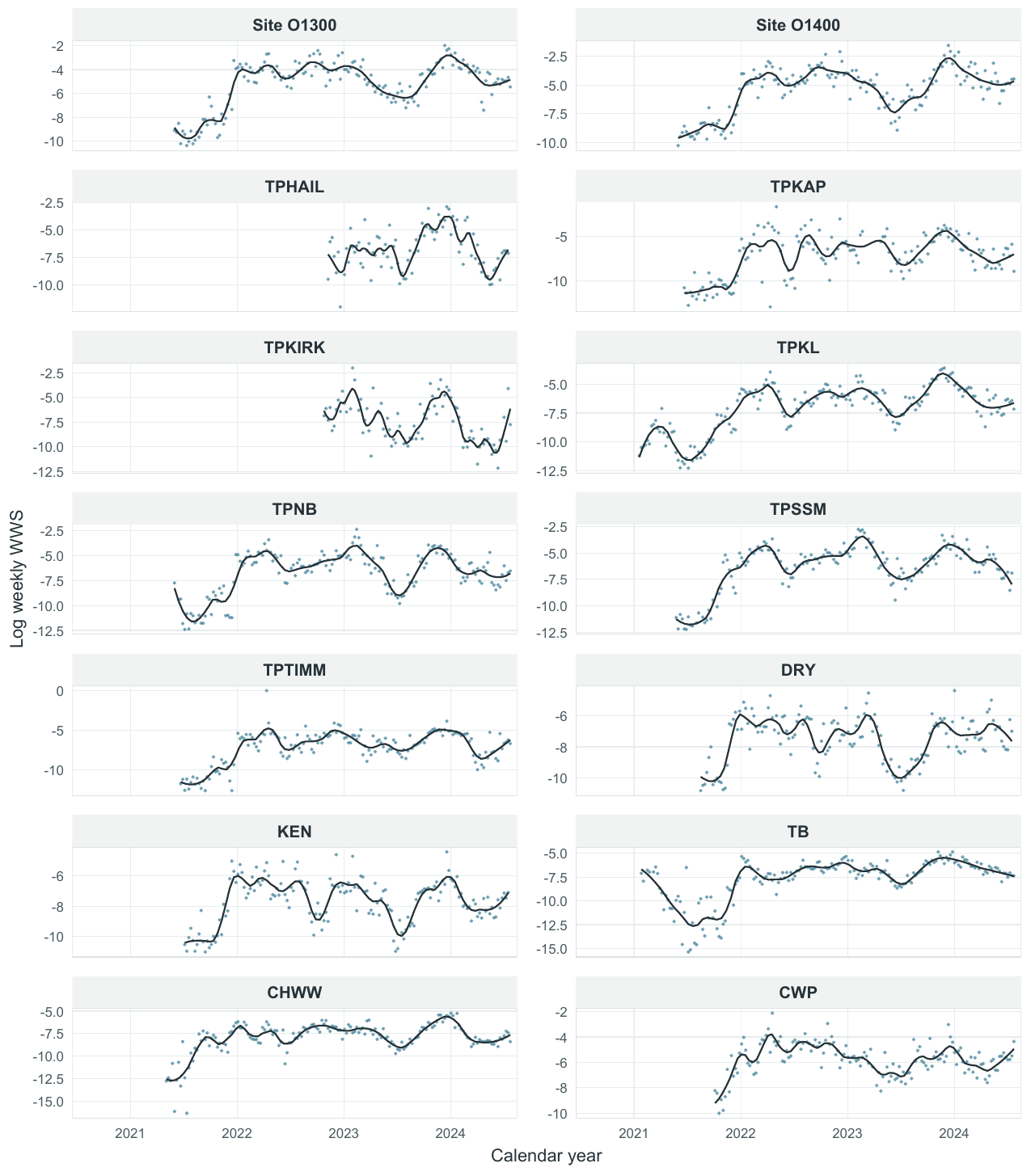}
\caption{Observed weekly log wastewater signals (part 3 of 4).}
\end{figure}

\begin{figure}[H]
\centering
\includegraphics[width=\textwidth,height=.77\textheight,keepaspectratio]{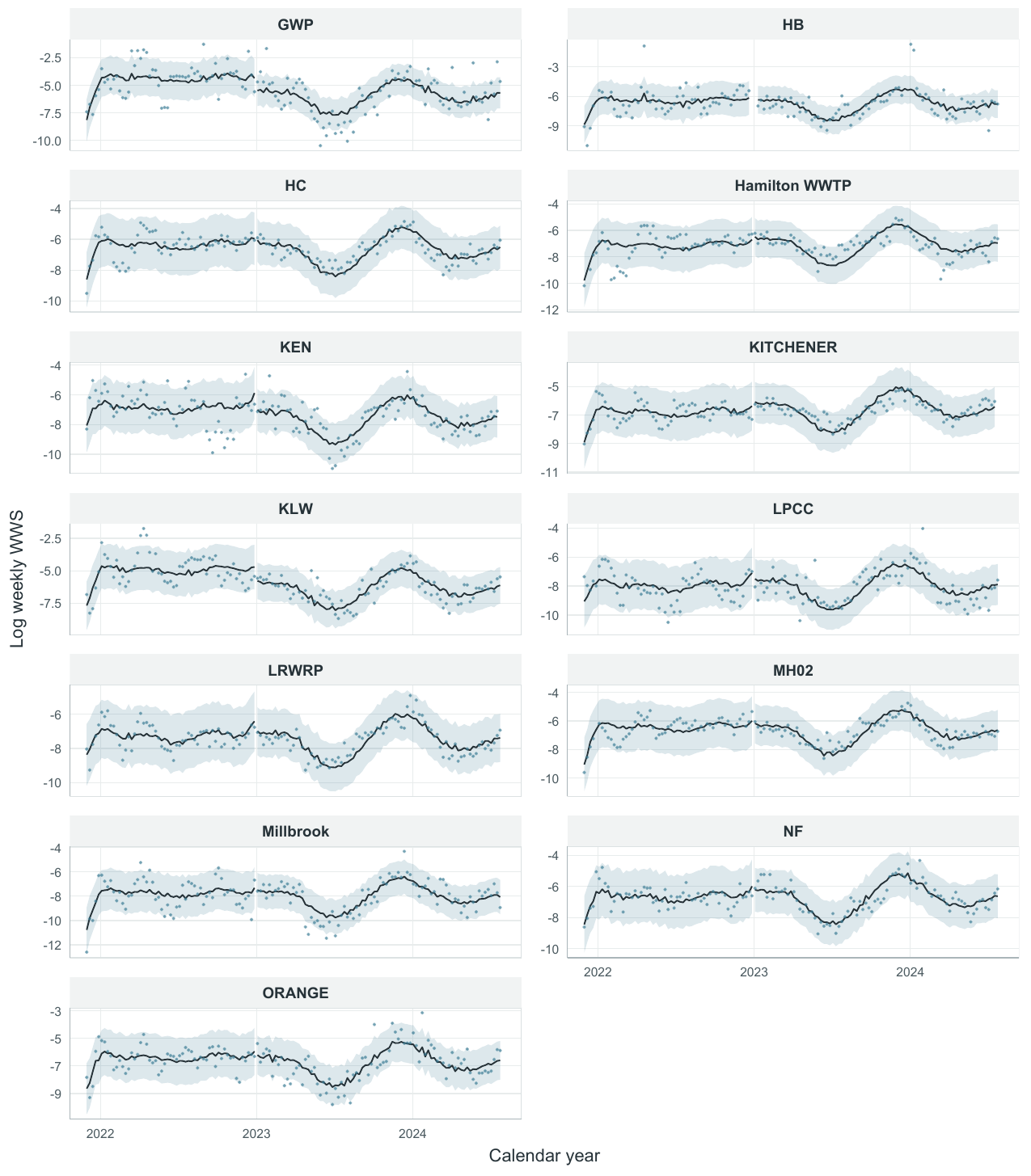}
\caption{Model~2 matched-period site fits (part 2 of 4).}
\end{figure}

\begin{figure}[H]
\centering
\includegraphics[width=\textwidth,height=.77\textheight,keepaspectratio]{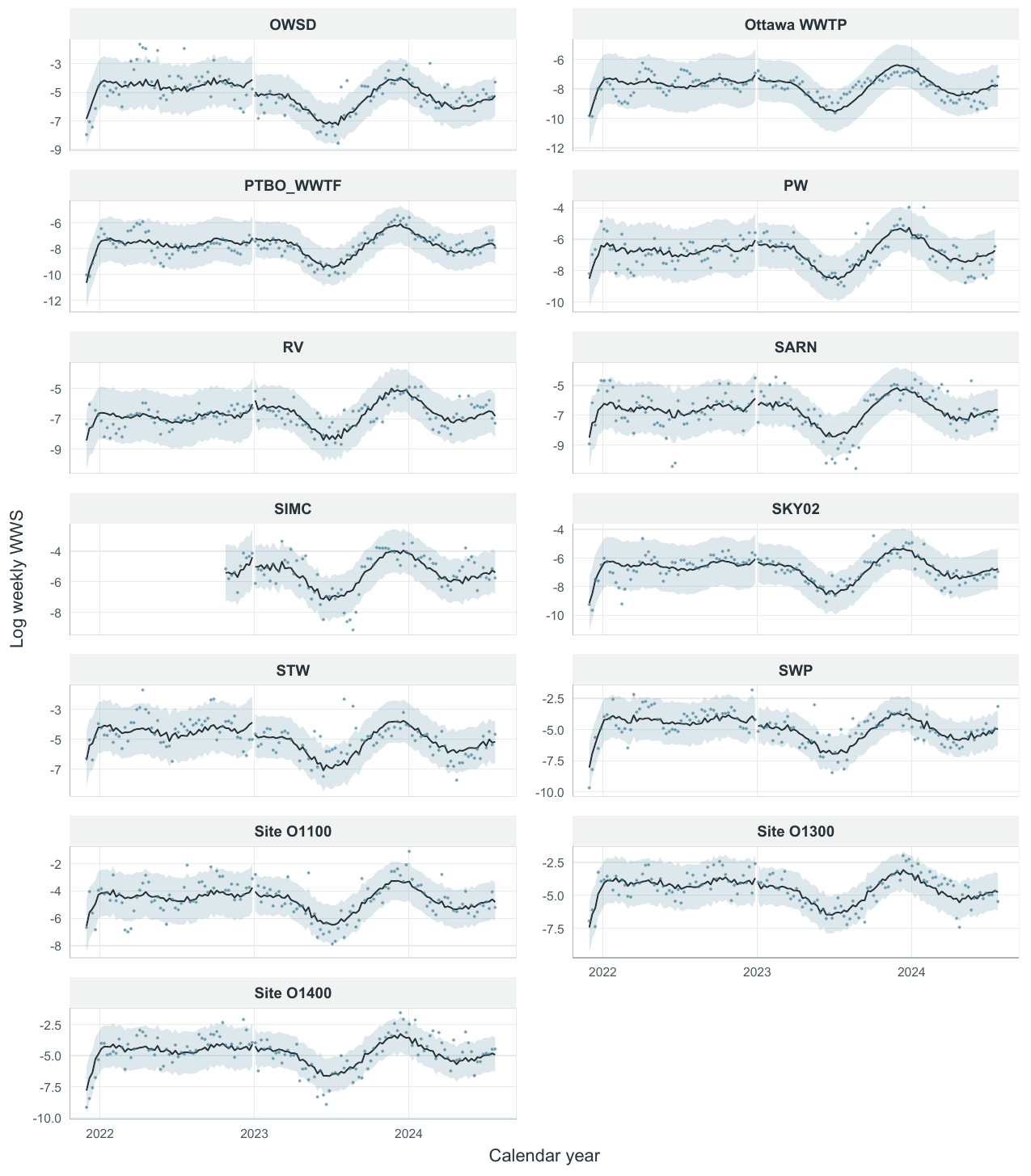}
\caption{Model~2 matched-period site fits (part 3 of 4).}
\end{figure}

